\documentclass[acmsmall,table]{acmart}

\AtBeginDocument{%
  \providecommand\BibTeX{{%
    \normalfont B\kern-0.5em{\scshape i\kern-0.25em b}\kern-0.8em\TeX}}}

\setcopyright{acmlicensed}
\copyrightyear{2023}
\acmYear{2023}
\acmDOI{XXXXXXX.XXXXXXX}

\acmJournal{PACMHCI}
\acmVolume{0}
\acmNumber{0}
\acmArticle{0}
\acmMonth{0}

\acmPrice{15.00}
\acmISBN{978-1-4503-XXXX-X/18/06}

\usepackage{xcolor}
\usepackage{subcaption}
\usepackage{tabularx}
\usepackage{multirow}

\usepackage{array}
\newcolumntype{L}[1]{>{\raggedright\let\newline\\\arraybackslash\hspace{0pt}}m{#1}}
\newcolumntype{C}[1]{>{\centering\let\newline\\\arraybackslash\hspace{0pt}}m{#1}}
\newcolumntype{R}[1]{>{\raggedleft\let\newline\\\arraybackslash\hspace{0pt}}m{#1}}

\begin{document}

\newcommand{\subsubsubsection}[1]{\textbf{#1.}}

\newcommand{\chm}[1]{\textcolor{red}{#1}}

\title[Personalizing Content Moderation on Social Media]{Personalizing Content Moderation on Social Media: User Perspectives on Moderation Choices, Interface Design, and Labor}

\author{Shagun Jhaver}
\email{shagun.jhaver@rutgers.edu}
\affiliation{%
  \institution{Rutgers University}
  \city{New Brunswick, NJ}
  \country{USA}
  }

\author{Alice Qian Zhang}
\email{zhan6698@umn.edu}
\affiliation{%
  \institution{University of Minnesota}
  \city{Minneapolis, MN}
  \country{USA}
  }

\author{Quanze Chen}
\email{cqz@cs.washington.edu}
\affiliation{%
  \institution{University of Washington}
  \city{Seattle, WA}
  \country{USA}
}

\author{Nikhila Natarajan}
\email{nn352@rutgers.edu}
\affiliation{%
  \institution{Rutgers University}
  \city{New Brunswick, NJ}
  \country{USA}
  }

\author{Ruotong Wang}
\email{ruotongw@cs.washington.edu}
\affiliation{%
  \institution{University of Washington}
  \city{Seattle, WA}
  \country{USA}
}

\author{Amy Zhang}
\email{axz@cs.uw.edu}
\affiliation{%
  \institution{University of Washington}
  \city{Seattle, WA}
  \country{USA}
}

\renewcommand{\shortauthors}{Jhaver et al.}

\begin{abstract}
Social media platforms moderate content for each user by incorporating the outputs of both platform-wide content moderation systems and, in some cases, user-configured personal moderation preferences. However, it is unclear (1) how end users perceive the choices and affordances of different kinds of personal content moderation tools, and (2) how the introduction of personalization impacts user perceptions of platforms' content moderation responsibilities. 
This paper investigates end users' perspectives on personal content moderation tools by conducting an interview study with a diverse sample of 24 active social media users. We probe interviewees' preferences using simulated personal moderation interfaces, including word filters, sliders for toxicity levels, and boolean toxicity toggles. We also examine the labor involved for users in choosing moderation settings and present users' attitudes about the roles and responsibilities of social media platforms and other stakeholders towards moderation. We discuss how our findings can inform design solutions to improve transparency and controllability in personal content moderation tools.
\end{abstract}



\maketitle
\section{Introduction}

\begin{figure*}
    \centering
    \begin{subfigure}[b]{0.675\textwidth}
        \centering
        \includegraphics[width=\textwidth]{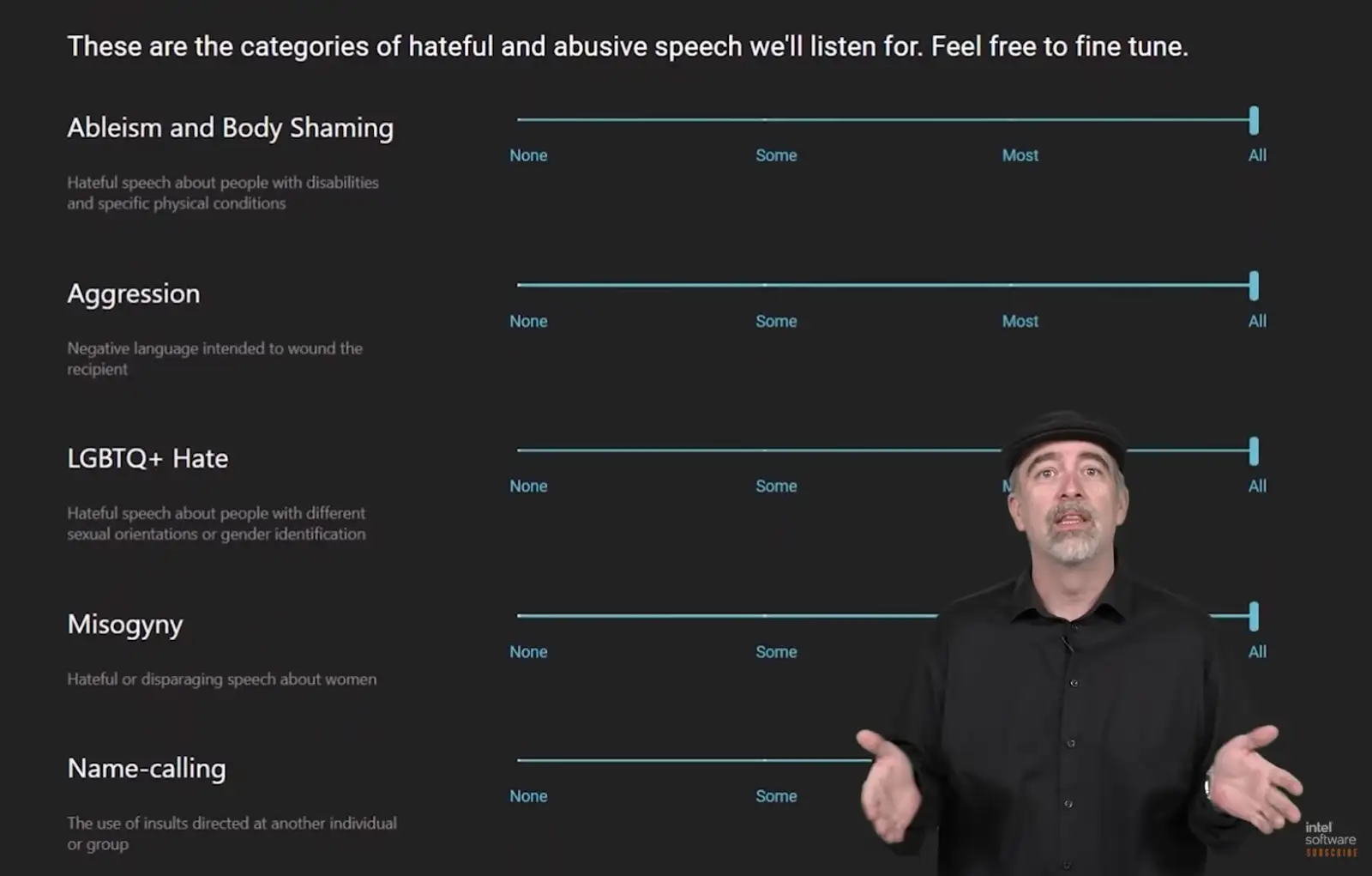}
        \label{fig:intel-one}
    \end{subfigure}
    \hfill
    \begin{subfigure}[b]{0.675\textwidth}   
        \centering 
        \includegraphics[width=\textwidth]{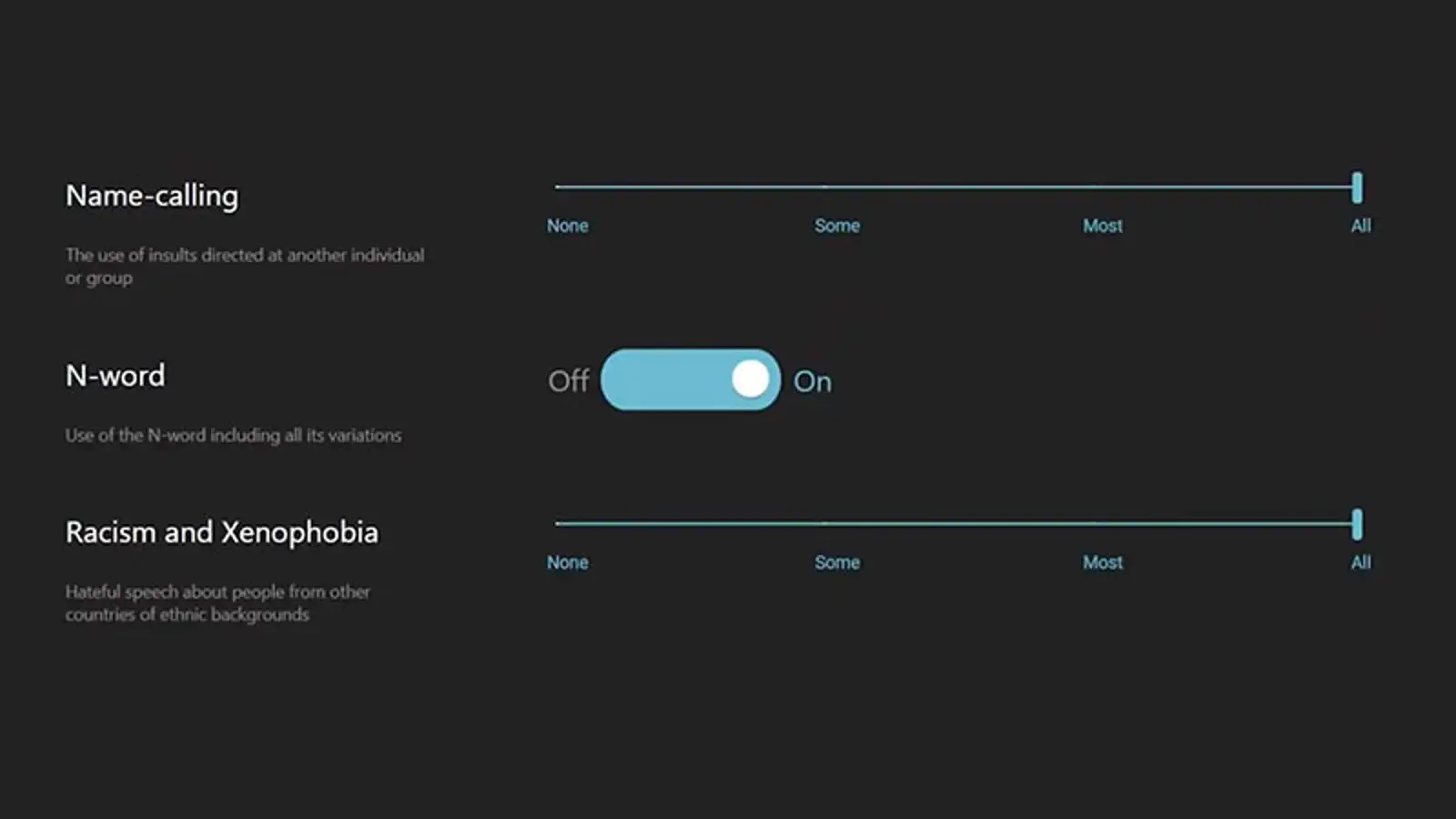}
        \label{fig:intel-two}
    \end{subfigure}
    \caption{Screenshots of Intel's implementation of a personal content moderation tool using sliders for different categories. During our interviews, we showed these images to our participants to elicit their suggestions.} 
    \label{fig:intel}
    \Description{Screenshots of Intel Moderation Interfaces}
    
\end{figure*}

At the Game Developers Conference in 2021, Intel unveiled Bleep, a new AI-powered tool for users to filter out online abuse in video game voice chat.\footnote{\url{https://youtu.be/97Qhj299zRM?t=1781}} As shown in screenshots of the demo in Fig. \ref{fig:intel}, the tool listed commonly encountered categories of abuse, like `Aggression' and `Misogyny,' each paired with a slider to control the quantity of each category that a user wants to hear \cite{porter_2021}.
Reactions were swift and heated across news and social media~\cite{polygon}.
One article objected to how the tool ``\textit{pitch[es] racism, xenophobia, and general toxicity as settings that can be tuned up and down as though they were...control sliders on a video game}''~\cite{vice}.
Other people argued that such a tool would be beneficial; for example, one Twitter user said, paraphrasing: \textit{``...as a trans woman gamer I pretty much always hear transphobia and misogyny online, but with this I can just bleep those words instead of muting people manually over and over...''}
Still others questioned the appropriateness of asking users to configure moderation settings, with a Twitter user saying: \textit{``...WTF kind of dystopic insanity is this control panel? Why would the onus be on the user to do the filtering?''}

The mix of responses highlights some of the difficult questions that arise when  tools for personal configuration of moderation settings are introduced on a social media platform. 
These tools have the potential to address online speech harms in a more user-initiated and personalized way.
However, the rich debate around the Intel demo shows that users have different opinions on and preferences for such tools.\newline

Since the mid-2000s, social media platforms' popularity has expanded dramatically  worldwide~\cite{twenge2019trends,union_2021,bahrini2019impact}. 
User-generated content (UGC) can be empowering, especially for those belonging to vulnerable or marginalized groups, as it allows users to dictate what content is created~\cite{reid2016children}.
However, the rules around what content is acceptable on a platform versus not continue to be narrowly shaped by the cultural norms of Silicon Valley, where most big platforms hosting UGC are located \cite{gillespie2018custodians}.
Given the normative differences across cultures~\cite{jiang2021understanding} and communities~\cite{weld2022makes}, including even those that are geographically co-located \cite{wellman2003social}, a one-size-fits-all solution to shaping online content would not be able to serve the disparate needs of millions of end users~\cite{jiang2022trade}.

\begin{figure}
    \centering
    \includegraphics[width=0.95\textwidth]{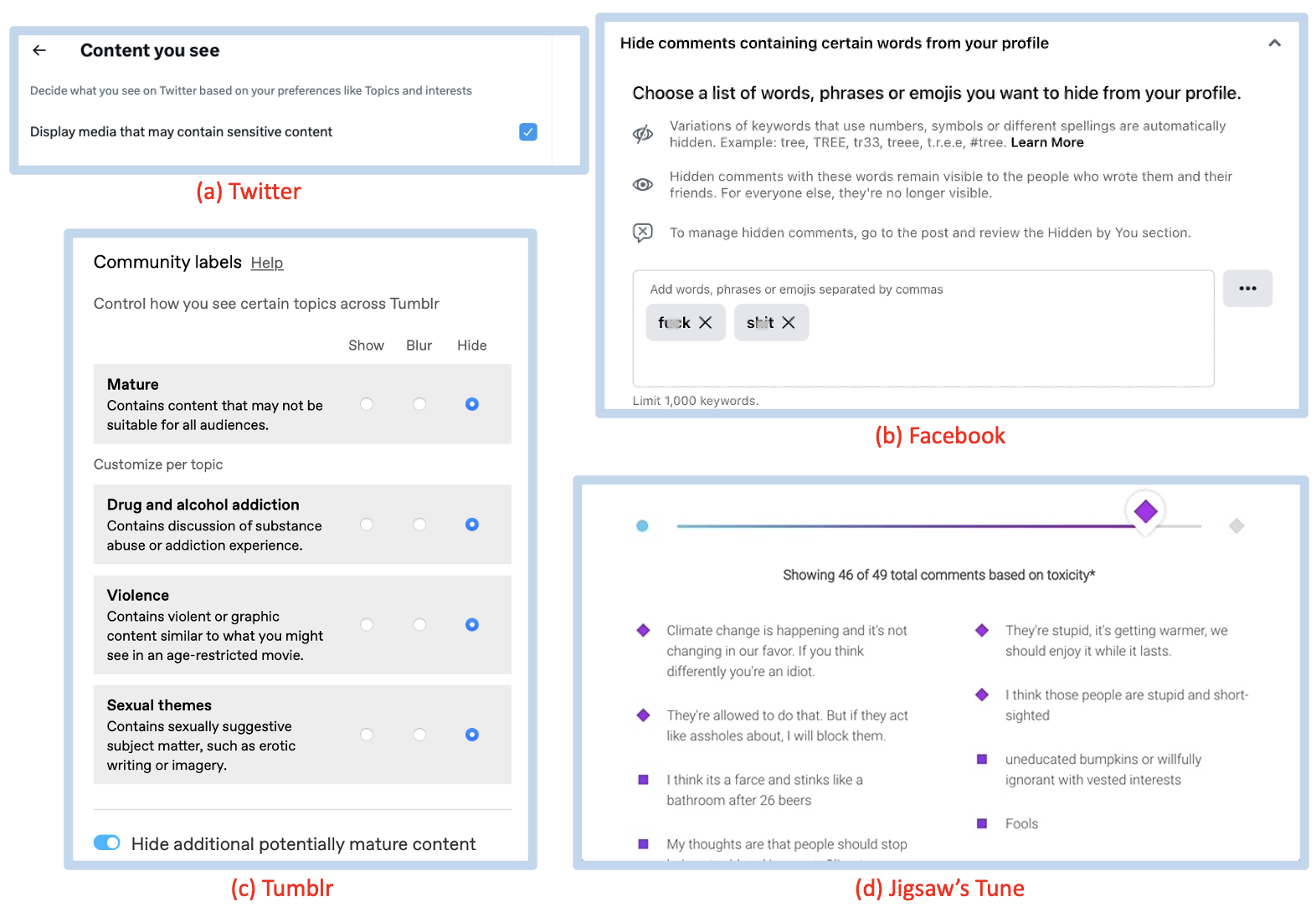}
    \caption{Examples of Personal Content Moderation Tools on (a) Twitter, (b) Facebook, (c) Tumblr, and (d) Jigsaw Tune.}
    \label{fig:examples_tools}
\end{figure}

Recognizing the inevitable conflicts regarding platform-wide content moderation, some industry leaders, scholars, and activists have called for an approach that we refer to as \textbf{personal content moderation}. We define personal content moderation as \textit{a form of moderation in which users can configure or customize some aspects of their moderation preferences based on the content of posts submitted by other users.} Recently, many platforms have begun to experiment with and offer such tools.
They are `personal' in that every user can configure them differently, and a user's configuration applies only to their own account. 
In addition, they are \textit{content-based} in that they help users configure moderation choices based on the characteristics of the content they encounter on the platform. 
Examples of tools for personal content moderation include toggles, sliders, or scales for `toxicity', `sensitivity', or other attributes, as well as word filter tools for filtering out user-specified phrases (see Fig. \ref{fig:intel}, \ref{fig:examples_tools}, \ref{fig:instagram}). 
These differ from more common \textit{account-based} personal moderation tools, such as being able to block or mute undesirable accounts individually or in bulk~\cite{jhaver2018blocklists,geiger2016,blockparty}.

In the context of internet history, personal content moderation tools, with all the promises and perils they entail, have found their moment. 
Critics and scholars are increasingly calling for mechanisms that move moderation decision-making away from centralized platforms and towards individual users to give them greater control over what they \textit{do not} want to see on social media \cite{narayanan2022Tiktok,leetaru_2019,engler_2022,masnick_2019}.
Popular platforms like Facebook, Instagram, TikTok, and Twitter now offer word filter tools \cite{jhaver2022filterbuddy} and sensitivity settings~\cite{instagram} for users to configure over their news feeds and over `Explore', `For You', and `Search' products.
In addition, emerging platforms and third-party personal moderation tools like Gobo Social \cite{bhargava2019gobo}, Bodyguard \cite{bastien_2021}, Block Party \cite{blockparty}, and Bluesky~\cite{bluesky} are letting users personalize their content moderation with even greater granularity.
Social media founders and owners such as Mark Zuckerberg and Jack Dorsey have advocated for personal content moderation settings that tune exposure to controversial content, such as nudity, violence, and profanity~ \cite{zuckerberg2021Building,jackdorsey}. 
A recent representative survey of 983 adult internet U.S. users documented the public appetite for personal content moderation by showing that 52.4\% of participants prefer relying on a personal moderation setting over the default platform-wide moderation to handle hate speech posts \cite{jhaver_zhang_2022}.

Despite growing interest in and adoption of personal content moderation tools, \textit{empirical} insights about  users' preferences for and attitudes towards such controls are scarce.
Instead, prior research on content moderation tools has primarily focused on tools used by people in moderator roles in a community-wide setting, such as volunteer moderation teams \cite{jhaver2019automated,matias2016civic,mcgillicuddy2016,seering2019moderator,wohn2019volunteer}. 
This article aims to fill this critical gap and uncover the prominent user preferences, concerns, and design considerations regarding personal content moderation. 

First, we sought to understand what considerations come into play when users have the ability to decide on personal moderation settings. 
A more restrictive configuration of settings could result in over-moderation, and a more lax configuration could lead to under-moderation.
Thus, we ask:

\begin{quote}
\textbf{RQ1:} \textit{What considerations do users have when deciding whether to be more or less restrictive in their choices for personal content moderation settings?}
\end{quote}

\noindent
Next, we wanted to understand users' challenges when interacting with and configuring personal content moderation tools as realized in different commonly-deployed designs, including toggles, word filters, and sliders. We therefore ask:

\begin{quote}
\textbf{RQ2:} \textit{
What issues do users encounter when trying to understand and control what happens when they interact with different interface designs for personal content moderation?
}
\end{quote}

\noindent
Finally, since the enactment of any voluntary moderation tool targeted at platform users is inextricably linked to the questions of labor involved (e.g., see \cite{dosono2019mod,roberts2019behind,steiger2021psy,wohn2019volunteer}), we raise the following research question:

\begin{quote}
\textbf{RQ3:} \textit{How do users perceive the labor in configuring these tools? How do they ascribe the distribution of responsibility for this labor between different stakeholders?}
\end{quote}


To gather deeper reflections and nuanced rationales for users' preferences, we conducted qualitative, semi-structured interviews with 24 social media users.
We considered that participants may not have interacted with personal content moderation tools on social media platforms.
Therefore, we developed a series of probes to prompt interviewees to consider different potential tool designs and elicit more informed opinions. 
To begin, we built a web application that simulates a social media feed of content along with a series of interactive controls that, when adjusted, re-configure the feed.
We preloaded this web application with data from a research dataset of tweets labeled with different toxicity levels~\cite{kumar2021designing}.
We implemented four types of controls based on personal  moderation tools that have been deployed or proposed in the past (Fig.~\ref{fig:interfaces}): a word filter, a toxicity toggle, an intensity slider, and a proportion slider. We describe these controls in detail in Sec. \ref{sec:methods}.

In the interviews, we asked participants to interact with all four control interfaces and inspect the resulting changes in the simulated feed while speaking aloud about their preferences. In addition, we showed them screenshots of personal content moderation controls developed in industry, including the Bleep tool and one offered by Instagram.
We used these probes to get more specific feedback on different design decisions.
These probes also
helped surface deeper reflections when we asked
interviewees about their perspectives on
whether and why they would (not) use such tools, moderation labor on the part of users, and platforms' responsibilities in the face of harmful content.

We found that while encountering offensive content on social media is a common experience, some prefer to just ignore such content, while others configure personal moderation settings, and still others get frustrated enough to quit social media altogether. 
Many interviewees were resistant to setting restrictive filters due to their fear of missing out on relevant posts and their desire to hear others out, even if the content might be offensive.
Our analysis also raises critical areas for improvement in the current designs of personal moderation tools from the perspective of end users: increasing clarity in the definitions of various interface elements, incorporating environmental/cultural context, offering appropriate levels of granularity, and a wider leveraging of example content as a means to provide transparency and enable greater control. Our findings also highlight users' understanding of the cognitive labor involved in personal moderation and, related to it, their perspectives on how platforms and lawmakers  share some responsibility.

We conclude by discussing the importance of addressing online harms while attending to users' desire to not overlook relevant content and how more context-aware personal content moderation tools can contribute to this goal. We highlight the value of clarifying the meaning of crucial interface elements and how doing so may necessitate an overhaul of current tools. 
We argue that designs that let users configure moderation exceptions for specified user groups or cultural contexts would increase the controllability of these tools.
We emphasize that configuring these settings should be iterative for users, and incorporating user preferences inferred from other interactions, such as reporting and seeking user feedback, could further improve their utility.
Finally, we examine platforms' roles and responsibilities in this space and how third-party developers and lawmakers can contribute to improving users' online experience.
\section{Related Work}
\subsection{Content Moderation}
\textit{Content moderation} is the organized governance of user-generated posts by information intermediaries and social media platforms to facilitate cooperation and prevent abuse on their sites \cite{grimmelmann2015virtues,roberts2019behind}. 
One of its main goals is to address online harms~\cite{jiang2022trade}. However, interpretations of online harm vary across cultures \cite{jiang2021understanding}, communities~\cite{weld2022makes}, and individuals \cite{jhaver2018view}, making it challenging to deploy one-size-fits-all moderation solutions.
We focus on content-based online harm that results from viewing undesirable content, such as hate speech or graphically violent images, on social media platforms.
Prior HCI research has contributed to our understanding of the diversity of online harms \cite{scheuerman2018safe,jhaver2018blocklists} and users' experiences of such harms \cite{scheuerman2018safe}. We add to this conversation by examining how end users personally grapple with the questions of trading off  viewing content that may be harmful to them against their fears of missing out on relevant content or desire to be open-minded.

In their early years, content on platforms like Facebook and YouTube was governed by relatively small review teams. 
Platform moderation rules and policies were also limited in scope \cite{klonick2017}.
As time passed and public pressure to remove offensive speech increased, platforms developed complex, sophisticated systems to aid their moderation functions (Table \ref{tab:mod-types}).
%
Many platforms now have automated site-wide filters that remove, in an initial pass, blatantly inappropriate posts, such as spam or child sexual abuse materials (CSAM) \cite{gillespie2018custodians}. Additionally, platforms deploy paid \cite{roberts2019behind} or volunteer \cite{mcgillicuddy2016} moderators to review and regulate the remaining posts. 
Many social media platforms are also feed-based, incorporating algorithms that sort posts based on users' prior interactions \cite{eslami2015always}.
 Each social media platform has developed various ad-hoc systems to implement these processes \cite{taylor2018watch,gillespie2018custodians}, yet each platform keeps the specifics of how it enacts its moderation decisions opaque \cite{jhaver2019automated}.
We add to this literature by surfacing social media users' perspectives on platforms' moderation apparatuses and what role, if any, users should play. We also consider how a greater emphasis on personal moderation might reconfigure how platforms conduct moderation more broadly.

\begin{footnotesize}
\sffamily
\begin{table}[]
\begin{tabular}{L{0.11\columnwidth}|L{0.15\columnwidth}|L{0.11\columnwidth}|L{0.1\columnwidth}|L{0.13\columnwidth}|L{0.11\columnwidth}|L{0.1\columnwidth}}
\toprule
\multirow{2}{*}{ }
 &
  \multicolumn{3}{c|}{\textbf{Account moderation}} &
  \multicolumn{3}{c}{\textbf{Content moderation}} \\ \cline{2-7}
 &
 \textbf{Actions} & \textbf{Purview} & \textbf{Impact} & \textbf{Actions} & \textbf{Purview} & \textbf{Impact} \\ \midrule
\textbf{Platform moderation} &
  ban or suspend users from the platform &
  all users on the platform &
  every user's view &
  remove content from the platform &
  all content on the platform &
  every user's view \\ \hline
\textbf{Community moderation} &
  ban users from a community &
  all users on the platform &
  every user's view &
  remove content from the community &
  content posted to the community &
  every user's view \\ \hline
\textbf{Personal moderation} &
  block a user from seeing one's content; mute a user from appearing in one's view &
  all users on the platform &
   blocked user's view; one's own view &
  \cellcolor{blue!25}remove content from one's view &
  \cellcolor{blue!25}content that appears in one's view &
  \cellcolor{blue!25}one's own view \\ \bottomrule
\end{tabular}
\caption{A characterization of different modes of content moderation. We have highlighted cells relevant to personal content moderation, this article's focus.}
\label{tab:mod-types}
\end{table}
\end{footnotesize}

Besides top-down moderation, platforms offer several technical mechanisms for users to shape what they see.
First, users primarily shape their feed by choosing to \textit{follow} (or \textit{unfollow}) or add as a \textit{friend} any account from which they wish to see more (or less). 
Second, users can express their perception of any post by one-click mechanisms such as \textit{like} or upvote/downvote. They can also report inappropriate posts using flagging tools, which trigger a post review by the site \cite{crawford2016}. Third, some feed settings can more broadly shape the news feed, such as the settings to change the look and feel of the site, account deletion or deactivation, and language or region settings \cite{hsu2020awareness}.

In addition to the above three mechanisms, users can also rely on \textit{personal moderation tools}. These tools let users configure their preferences for the activity they want to \textit{avoid}. 
Personal moderation tools fall into two categories: account-based and content-based. \textit{Personal account moderation tools} let users mute or block \textit{an account} to prohibit future interaction with it. Some third-party developers have built upon this functionality to enable mass blocking \cite{geiger2016,blockparty} or peer-assisted blocking~\cite{mahar2018squadbox}, and researchers have examined the utility and deficiencies of such tools \cite{jhaver2018blocklists,geiger2016}. On the other hand, \textit{personal content moderation tools} let users make moderation decisions \textit{on each post} based on its content alone and regardless of its source. Personal content moderation tools include tools to mute specific keywords \cite{jhaver2022filterbuddy}, remove\footnote{Note that in the context of personal moderation tools, post removals occur only for the configuring account. Others users can still see the removed content.} NSFW (not safe for work) content, and set up sensitivity controls (see Fig. \ref{fig:examples_tools} and \ref{fig:instagram} for examples).

In our work, we focus specifically on personal content moderation tools.
We choose this focus because it has become increasingly common that the posts shown in users' feeds are not made by people they follow---an example is TikTok's `For You' page, which is typically dominated by stranger accounts. 
The same can be said for users who use the search functionality on platforms such as Instagram or Pinterest to find content.
In these scenarios, personal account moderation tools may not have as much impact on what users see due to the high proportion of novel accounts. This issue may also crop up in specific contexts, such as some gaming platforms where users frequently communicate with strangers \cite{kou2021flag}.
As a result, personal controls that enable users to configure moderation based on content become more important. Indeed, in recent years, platforms have begun offering more personal content moderation tools~\cite{instagram}, and industry leaders have called for more user controls for moderation~\cite{zuckerberg2021Building,jackdorsey}, leading to experimental efforts like Intel's Bleep (Fig. \ref{fig:intel}) and Jigsaw's Tune (Fig. \ref{fig:examples_tools}(d)) tools.

As far as we know, this is the first paper that conceptually identifies personal content moderation as a distinct category and investigates user preferences for them.
Prior research that focuses specifically on personal content moderation tools is scarce. One related work is Jhaver et al.'s research on need-finding and design exploration around word filter tools \cite{jhaver2022filterbuddy}. However, they focus on word filters used by content creators to delete comments on their content for all viewers, not users muting content in their personal feed. 
We note that specialized categories of users---such as advertisers, admins, and community moderators---have a broader range of moderation tools available to them, such as Automod \cite{jhaver2019automated} or Sentropy \cite{hatmaker_2021}.
Prior research on designing such tools identifies some challenges relevant to our work, e.g., rule-based configuration producing false positives \cite{jhaver2019automated,chandrasekharan2019crossmod}.
However, our focus here is on moderation tools available to general users that do not rely on collaborative moderation teams and that only affect each user's own view. 







\subsection{User Control and Transparency in Interactive Recommender Systems}
This section briefly reviews prior literature on interactive recommender systems as it offers valuable insights for designing personal moderation tools.
Recommender systems are ubiquitously deployed today to solve the problem of information overload and to increase engagement \cite{jannach2016user,jin2017different}.
Over the past decades, extensive research has been conducted to develop and deploy algorithms that suggest relevant items to a user \cite{adomavicius2005toward}.
However, several challenges still need to be solved that prohibit recommender systems from realizing their full potential \cite{he2016interactive}.

There is a growing awareness that the effectiveness of recommender systems goes beyond recommendation accuracy \cite{pu2012evaluating,konstan2012recommender,burrell2019users}. Research on integrating human values such as diversity, serendipity, and trust \cite{jin2017different,lee2018understanding,stray2022building} into recommendations is gaining interest.
As part of this push into user values, which may differ depending on the user,  user controls like toggles and sliders have been proposed for interacting with recommender systems. This integration supports personalization, transparency, and controllability of the recommendation process~\cite{he2016interactive}.
These techniques form \textit{interactive recommender systems} \cite{he2016interactive,mahmood2007learning}, and the personal content moderation tools we study in this work offer similar controls. 
Advanced interactive recommender systems also incorporate contextual information (e.g., location, current activity, interest) to generate recommendations that tailor to the current needs of the user \cite{adomavicius2011context}. We examine the importance of incorporating relevant context in personal moderation tools from end users' perspectives.

In general, recommender and content moderation systems are similar in that they both act on the content one sees, where tuning a recommendation algorithm to reduce a specific type of content may have a de facto effect akin to moderating it out~\cite{gillespie2022not}. When done by platforms, this has often been described critically as ``shadowbanning''~\cite{vice2}.
However, a significant point of differentiation between the two is the goals behind them. Recommender systems aim to shape what content is shown to the user to make a better recommendation \cite{jannach2016user}. In contrast,  moderation systems focus on removing content types that are likely to harm the user. 
As a result, users may have very different ideas about how personal controls for recommendation versus moderation should be designed.
For instance, a user seeking to remove certain harmful content might desire more precise controls to be more confident they will not encounter that content compared to a recommendation control.
Indeed, in major platforms, recommendation and moderation often have different policies and separate teams dedicated to them.

Still, there are some aspects that users may find desirable in the design of both kinds of controls. Thus, we consider what we can learn from the interactive recommender literature.
For instance, interactive recommender systems aim to provide \textit{transparency} into the black-box nature of a recommendation system by explaining its inner logic to end users \cite{sinha2002role,vig2009tagsplanations}.
Exposing the reasoning and data behind a recommendation may help increase users' confidence in that recommendation and improve their acceptance of the system \cite{abdul2000supporting,herlocker2000explaining}. 
We examine how to provide transparency in the context of content-based personal moderation tools, where there is also an inner logic to the filter implementation.

Second, a related objective of interactive recommender systems is to offer \textit{justification}, which refers to describing \textit{why} the user gets a specific recommendation rather than \textit{how} that recommendation was selected or how the system works \cite{tintarev2011designing,vig2009tagsplanations}.
In some cases, justification may be preferred over transparency as descriptions of the underlying recommendation technique may be too complex or cannot be revealed to protect intellectual property \cite{herlocker2000explaining,tintarev2011designing,vig2009tagsplanations}.
We take inspiration from this literature by examining whether offering examples can help with understanding and if examples improve user acceptance of such tools.

The third key objective of interactive recommender systems is to improve \textit{controllability}, or how much the system supports users in fine-tuning their configurations.
Controllability can compensate for deficiencies in recommendation algorithms by incorporating user input and feedback to tailor recommendations to users' changing preferences \cite{he2016interactive}.
Prior research has shown that in some application domains, users appreciate being more actively involved in the process and in control of their recommendations \cite{knijnenburg2012inspectability,dooms2014improving,parra2015user,vaccaro2018illusion,Roy2019AutomationAI}.
In the same vein, lower levels of user control can negatively influence the perceived quality of recommendations \cite{harper2015putting}. On the other hand, too many control interfaces can make user interfaces challenging to understand \cite{andjelkovic2016moodplay} and increase user cognitive load \cite{jin2017different}.
We examine users' perspectives on controllability in personal moderation tools by comparing interfaces that offer different granularities of control.

In summary, we take a user-centered approach in this work to examine how the types of personal moderation interfaces that commonly appear on mainstream social media platforms serve the preceding objectives of transparency, justification, and controllability. 




\subsection{Responsibilities of Platforms and Lawmakers}
With the dramatic migration of online discourse to social media platforms in recent years, questions about how they structure social activity and the rights and obligations they have for the speech they facilitate are becoming increasingly important \cite{denardis2015internet,mackinnon2015fostering,obar2015social,van2013culture,wagner2013governing}.
In the United States, Section 230 of the U.S. Communications Decency Act (CDA) protects intermediaries (including online platforms) from liability for illegal content sharing by users while also protecting them from liability when they do remove content that violates company policy \cite{goldman2018complicated}.
This law reflects a fundamental reluctance to constrain speech inside the U.S.
Most countries in the European Union and South America also do not hold platforms liable for their users' illegal posts as long as they comply with state requests to remove such posts \cite{mackinnon2015fostering}.

In practice, however, nearly all platforms go beyond the legal requirements for removing inappropriate content, and user experience is shaped much more by platform policies than by legal restrictions. Critiques about these policies often mirror longstanding debates about the character of acceptable public discourse.
Gillespie highlights examples of platform governance controversies that reflect society's broader public discourse concerns: 

\begin{quote}
    \textit{``which representations of sexuality are empowering and which are explicit, and according to whose judgment; what is newsworthy and what is gruesome, and who draws the line; how do we balance freedom of speech with the values of the community, with the safety of individuals, with the aspirations of art, and with the wants of commerce''} \cite{gillespie2017governance}.
\end{quote}

Gillespie argues that we should not leave the responsibility of resolving these fundamental tensions of social and public life to platforms alone but instead govern collectively as citizens \cite{gillespie2018custodians}. Personal moderation tools empower end-users to specify their boundaries of acceptable conversation.
In contrast, if we rely on platforms alone to serve acceptable public discourse for everyone, policies that enforce the most sanitized content, such as only what is appropriate for underage audiences or the content that Silicon Valley workers find acceptable, may prevail.
Such policies may result in unnecessary censorship of otherwise contextually appropriate discourse~\cite{york2022silicon}.

Many legal and media scholars have theoretically examined the dynamics of free expression online \cite{balkin2013old,godwin2003cyber,lessig2009code,litman1999electronic}. However, there needs to be a more empirical understanding of how end users perceive their role in content moderation systems and where they might draw the line for themselves if given the option.
Adding personal moderation complicates the role of platforms and policymakers in striking a balance between regulating inappropriate posts and protecting free speech, as platforms could leave up more borderline content and let users decide whether they want to see it.
However, this raises critical questions about the physical and emotional labor involved in enacting moderation and the ethics of exploiting the unpaid work of end users to improve platform offerings \cite{dosono2019mod,roberts2019behind,steiger2021psy,wohn2019volunteer,jhaver2019automated}. 
It also raises ethical questions about what speech users should get a say in versus what speech platforms should be compelled to take down for all users.
We leverage our participants' examination of a sample of personal moderation interfaces to induce them to reflect on the questions of labor and the responsibilities of different stakeholders in this space.



%


\section{Method} \label{sec:methods}
Our university's IRB\footnote{We will disclose the University name after the peer review process is completed.} approved this study. Our study involved 24 semi-structured interviews with active social media users. During the interviews, we used a technology probe~\cite{hutchinson2003technology}---in this case, an interactive, toy social media feed that simulated how four personal moderation interfaces would work---along with additional screenshots of existing tools to prompt each participant in order to answer RQ2. Every participant interacted with the four interfaces. Additionally, we asked participants questions about the contexts in which they could use these tools, the tradeoffs they consider in using them, and their perspectives on the labor involved to answer RQ1 and RQ3. We now describe the details of our study design.

\subsection{Simulating Personal Content Moderation Interfaces}
We began by systematically observing personal content moderation tools and interfaces on the most popular social network platforms available in English.\footnote{We focused on the English language platforms with at least 100 million active users available on \url{https://en.wikipedia.org/wiki/List_of_social_platforms_with_at_least_100_million_active_users}} These platforms included Facebook, Twitter, Instagram, and TikTok, among others. 
We created a new account or used our existing account on each site and looked into its settings page to observe the options available for user-enacted moderation. 
We also examined the options available through third-party moderation tools like Gobo Social \cite{bhargava2019gobo}, Bodyguard \cite{bastien_2021}, and Intel's Bleep tool. 
Given our focus on personal \textit{content} moderation tools, we did not consider account-based settings such as muting and blocking accounts.
We focused only on tools available to end users engaged in consuming content rather than on community moderator tools, which often deploy more specialized automated moderation tools~\cite{jhaver2019automated}, or content creator tools that moderate comments on their content~\cite{jhaver2022filterbuddy}. 
Through these observations, we concluded that the commonly deployed moderation interfaces included:

\begin{enumerate}
    \item \textit{Toggles}. These tools offer a `yes/no' choice so users can allow or avoid seeing potentially undesirable content. For example, Twitter offers on its settings page a checkbox for ``Display media that may contain sensitive content'', and TikTok has a `Restricted mode' that may be turned on to restrict `inappropriate' videos. Tumblr and Bluesky have toggles for different categories of content, as well as a `blur' or `warn' option in addition to showing or hiding.
    \item \textit{Word filters}. These tools let users configure a list of phrases; posts containing any of those phrases are automatically removed or moved to a separate folder for human review. Word filters are commonly available on most popular platforms for muting content on one's feed, including Twitter, TikTok, and Instagram.
    \item \textit{Sliders}. These tools let users specify on a 3--5 point scale their desired thresholds for a given concept over the posts they see. 
    Instagram has a 3-point scale for `sensitivity' but presents options in a list with radio buttons (Fig.~\ref{fig:instagram}). Intel's Bleep has 4-point sliders for a number of categories. Most versions of sliders have levels according to proportion of moderation, but Gobo Social's levels are according to the intensity of the concept.
\end{enumerate}
\begin{figure*}
    \centering
    \begin{subfigure}[b]{0.475\textwidth}
        \centering
        \includegraphics[width=\textwidth]{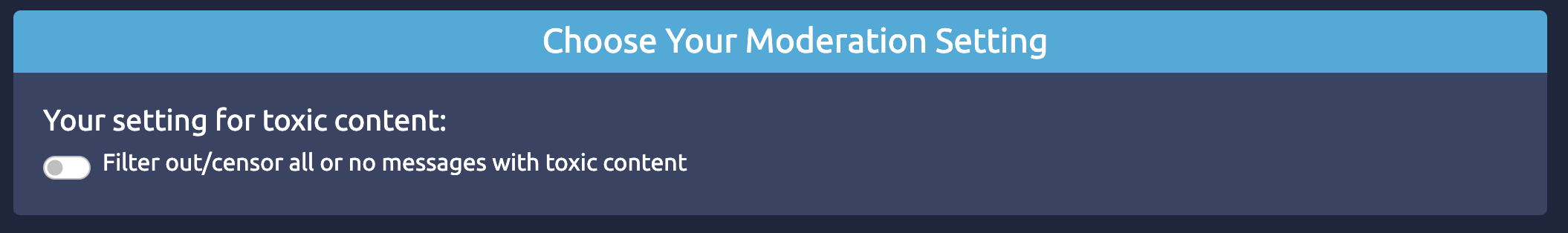}
        \caption {Binary Toggle}    
        \label{fig:binary}
    \end{subfigure}
    \hfill
    \begin{subfigure}[b]{0.475\textwidth}   
        \centering 
        \includegraphics[width=\textwidth]{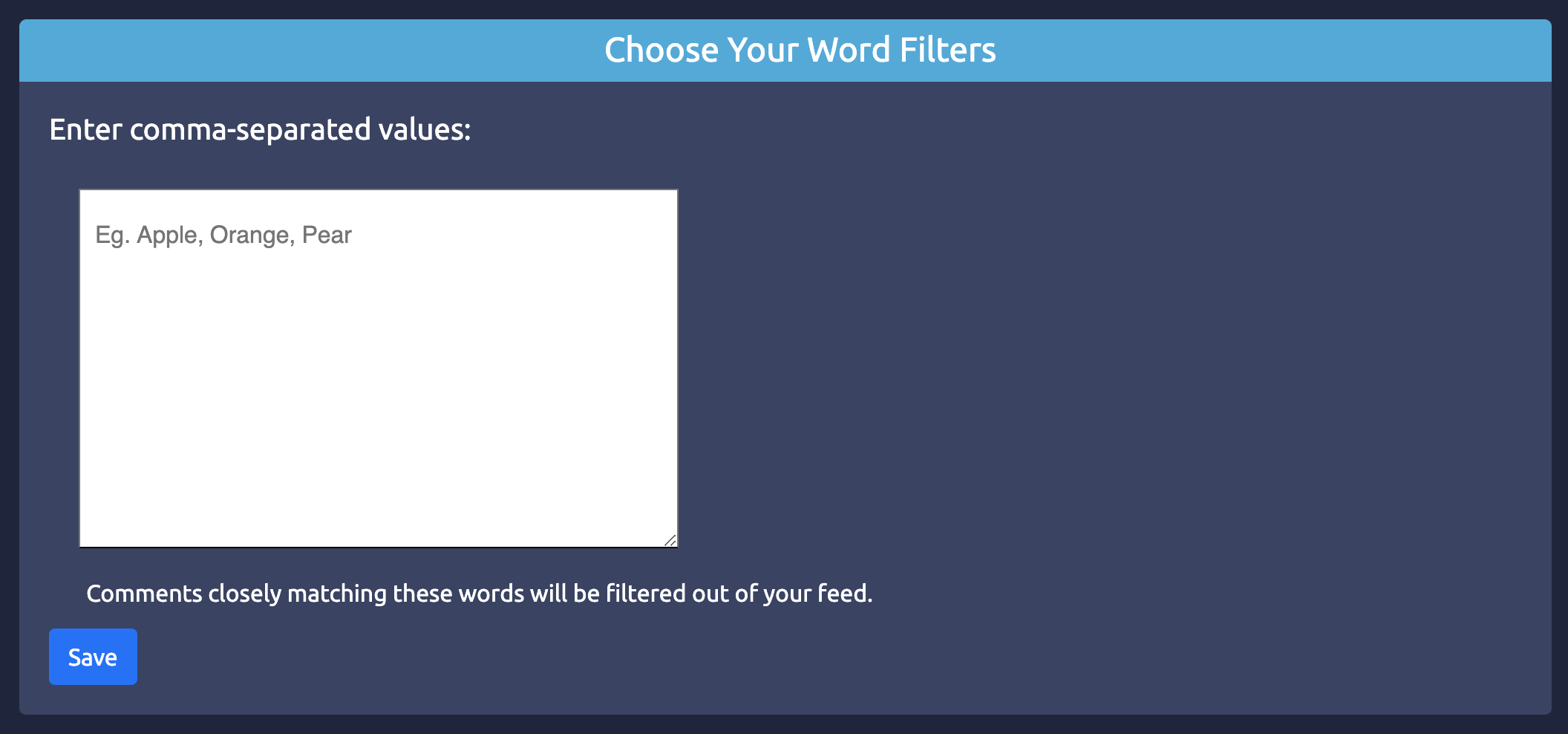}
        \caption {Word Filter}   
        \label{fig:word_filter}
    \end{subfigure}    
    \vskip\baselineskip
    \begin{subfigure}[b]{0.675\textwidth}  
        \centering 
        \includegraphics[width=\textwidth]{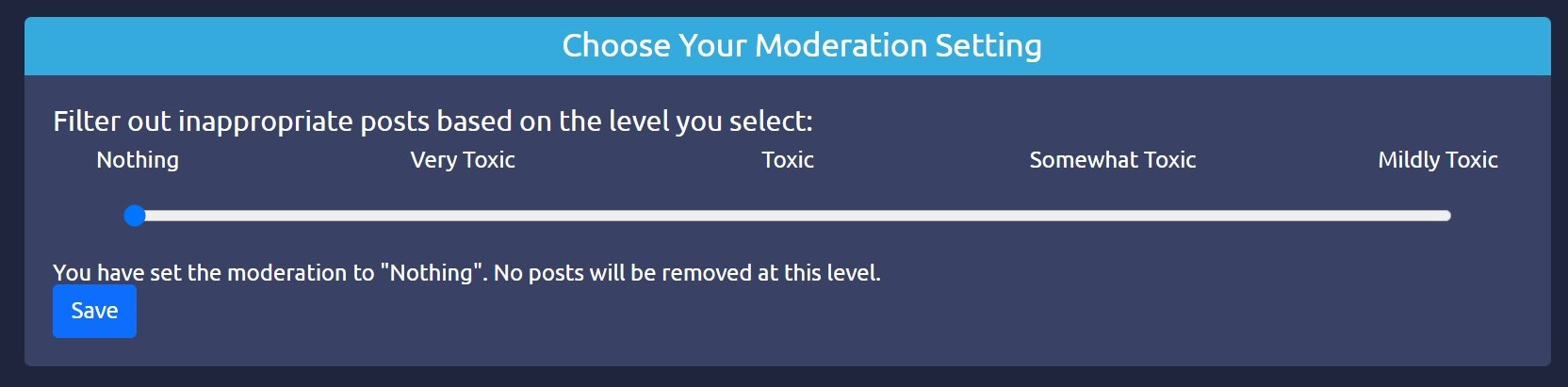}
        \caption {Intensity Slider}    
        \label{fig:intensity}
    \end{subfigure}
    \hfill
    \begin{subfigure}[b]{0.675\textwidth}   
        \centering 
        \includegraphics[width=\textwidth]{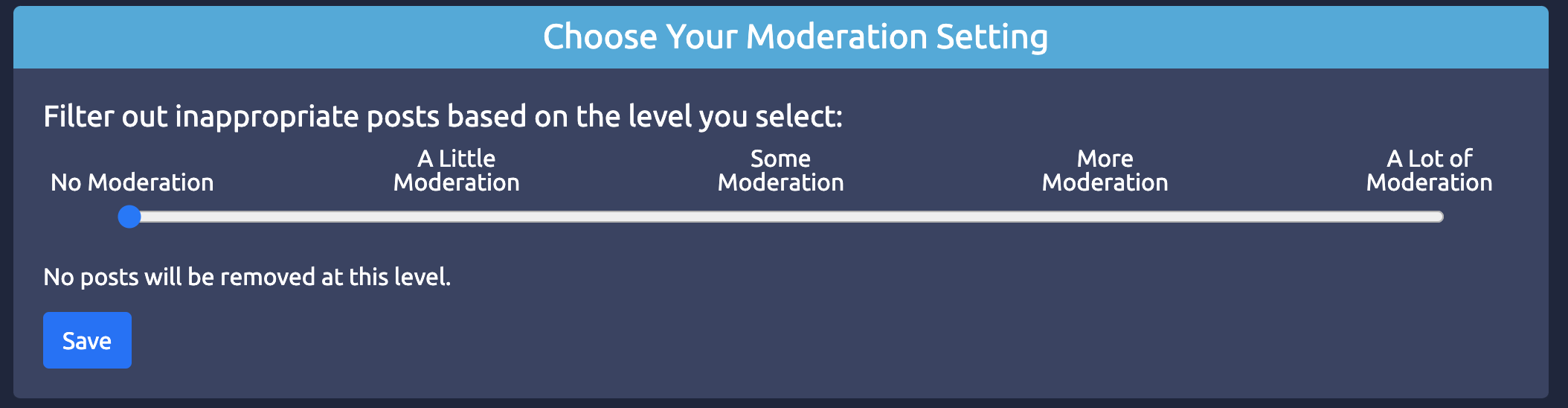}
        \caption {Proportion Slider}   
        \label{fig:proportion}
    \end{subfigure}
    \caption{Implementations of four moderation interfaces used in our study. (a) Binary toxicity filters. (b) Word filters. (c) Intensity sliders. and (d) Proportion sliders. These interfaces were inspired by personal moderation tools commonly available on popular social media platforms.} 
    \label{fig:interfaces}
    \Description{Interfaces of four moderation implementations deployed in this study.}
    
\end{figure*}

We created a website to simulate a user's experience viewing a news feed of comments on a social media platform much like Twitter. Based on our observations of popular moderation interfaces described above, we also built four moderation interfaces on this website that let users change their personal moderation settings for each interface and observe a changed news feed. These interfaces included a toggle for `toxicity', a word filter, a 5-level slider for `toxicity' that has levels labeled according to intensity or degree of toxicity, and a 5-level slider for `toxicity' that has levels labeled according to proportion or amount of moderation; see Fig. \ref{fig:interfaces} (a-d).
We built and used these interfaces as interview probes for eliciting user opinions on the design space of personal content moderation tools.
The intensity and proportion sliders were set to the default option of ``no moderation.''

We focused on these interfaces because they afforded different levels of granularity, transparency, and control in their design---themes we wanted to explore in our user study. 
The set of interfaces we developed is not an exhaustive representation of the currently available moderation tools. 
Further, some available tools, e.g., Intel's Bleep or Tumblr's labels, have a more detailed granularity of moderation settings.
We selected and built these interfaces to demonstrate to interviewees ways in which the design of personal content moderation tools could vary. 
We also sought participants' views on additional tools like Intel's Bleep and Instagram's sensitivity filters to aid this design exploration further.

We implemented our interfaces for only text comments as labeled datasets for toxicity were more readily available for textual rather than multimodal content. 
Still, text-only comments form the bulk of content on many platforms. We also asked our participants' views on multimodal personal content moderation tools offered by Intel and Instagram. 

\subsubsection{Comment Curation}
To curate comments for our simulation website, we began with a labeled dataset of 107,620 comments from Twitter, Reddit, and 4chan \cite{kumar2021designing}. 
We filtered for only the Twitter comments because comments from other platforms were more difficult to understand out of context. Moreover, these platforms' thread structure differs from the Twitter-like feed we  simulated in our study. 
We used the following attributes available for each comment in this labeled dataset: (1) its text; (2) whether the comment was profane, a threat, an identity attack, an insult, or sexual harassment; and (3) five independent ratings for each comment on a five-point scale from 0-4 ranging from ``Not at all toxic'' to ``Extremely toxic.'' 

Next, we filtered only those comments for which raters in the dataset had a consensus on the toxicity levels. To achieve this, we excluded all comments where the difference between the highest and lowest rating was greater than one point. Next, we calculated the average of the five toxicity ratings for each comment and bucketed all comments into five windows of equidistant toxicity averages. We call these buckets B1, B2, ..., B5, with B1 representing the bucket with the lowest average toxicity and B5 the highest. From each bucket B2--B5, we selected a random sample of 20 comments. 

All coauthors then manually reviewed these 80 comments and selected 5 comments from each bucket that were comprehensible without additional context. We selected comments that, taken together, represented topical diversity (e.g., cooking, food, sports, politics, celebrity news) and diversity of inappropriate behaviors (e.g., profanity, threats, identity attacks, insults, sexual harassment). We achieved this through mutual discussions and clarifications. We also selected an additional random sample of 30 comments from bucket B1.

\subsubsection{Interface Implementation}
We simulated our news feed to show 30 comments by default: 5 comments each from buckets B2--B5 and 10 from B1, randomly shuffled together. We implemented each moderation interface as follows:

\begin{enumerate}
    \item \textit{Binary toxicity toggle.} By default, the toggle setting was turned off. When the toggle setting was turned on, we removed comments from buckets B2-B5 and showed only the 30 comments from bucket B1. 
    \item \textit{Word filters.} We excluded all comments matching the keywords configured by the user in their word filters. We replaced the excluded comments with additional bucket B1 comments to maintain the total number of comments at 30.
    \item \textit{Intensity slider.} This slider removed progressively more toxic comments as the slider level moved from `Mildly Toxic' to `Very Toxic'. For example, when users moved the slider level from `Nothing' (default level) to `Very Toxic', we removed five comments from bucket B5. We replaced them with 5 randomly selected bucket B1 comments to maintain the total number of comments at 30.
    \item \textit{Proportion slider.} This slider's implementation pre-classified each comment as non-toxic or toxic, depending on whether the comment was in bucket B1 or not, respectively. When moved to its immediate right, each slider level removed 5 randomly selected toxic comments from the feed and replaced them with five bucket B1 comments.
\end{enumerate}

Our website lets users choose one of these four moderation interfaces at a time. Once a new interface is selected, the settings for all other interfaces are programmed to reset to their default levels. For example, if a user set up a few keywords in word filters and then used the intensity slider interface, our implementation would not filter out the keywords configured in word filters.
This setup resulted in the absence of interaction effects, which made it easier for users to understand the operations of each new interface they used.

\subsection{Participant Recruitment}
We recruited interview participants for our study by advertising on social media platforms such as Facebook, Twitter, and Reddit. Our calls for participation asked candidates to submit an online form that helped us obtain relevant information about the candidates. This form
included questions about whether the participants used social media daily, which social media platforms they used, whether they encountered toxic content on social media, and their demographic information. We also included an open-ended question: ``What is your perspective on how social media platforms should deal with toxic content?''

Analyzing the responses to this form, we selected candidates who seemed to have some familiarity with or be more likely to benefit from personal moderation tools, given their experiences, since our focus was on how best to get insights about their use and potential improvements.
We examined the depth and clarity of users' responses to our open-ended question and their previous encounters with toxic content to guide our participant selection.
We also oversampled individuals from marginalized groups, such as Black and LGBT users, since such users are more likely to experience online harm \cite{schoenebeck2020drawing} and, therefore, could benefit from more advanced moderation tools. Further, we ensured that our interview sample represented diverse ages, occupations, and countries.
Table \ref{table:participants} shows a  list of the participants.

\begin{table}[]
\small
\begin{tabular}{|l|llll|}
 \hline
\textbf{\#} & \textbf{Age} & \textbf{Gender} & \textbf{Occupation}              & \textbf{Country} \\ \hline 
P1          &      27        & Female          &           Grad student           & USA              \\ \hline
P2          & 24           & Female          & Grad student                     & USA              \\ \hline
P3          & 23           & Female          & Grad student                     & USA              \\ \hline
P4          & 31           & Male            & Grad student                     & USA              \\ \hline
P5          & 34           & Male            & Journalist                       & UAE              \\ \hline
P6          & 26           & Male            & Grad student                     & USA              \\ \hline
P7          & 21           & Female          & Student                          & USA              \\ \hline
P8          & 26           & Female          & Grad student                     & USA              \\ \hline
P9          & 25           & Male            & Grad student                     & USA              \\ \hline
P10         & 40           & Female          & Stand-up Comedian                & India            \\ \hline
P11         & 40           & Male            & Angel Investor                   & India            \\ \hline
P12         & 56           & Male            & Editor                           & India            \\ \hline
P13         & 23           & Female          & Engineer                         & USA              \\ \hline
P14         & 44           & Male            & Applications Manager             & The Netherlands  \\ \hline
P15         & 48           & Male            & Documentation Specialist         & Belgium          \\ \hline
P16         & 34           & Male            & Civil Engineer                   & UK               \\ \hline
P17         & 26           & Male            & Journalist                       & UK               \\ \hline
P18         & 29           & Female          & Physical Therapist               & USA              \\ \hline
P19         & 39           & Male            & Psychologist                     & Canada           \\ \hline
P20         & 33           & Male            & Director of Content \& Marketing & USA              \\ \hline
P21         & 23           & Male            & Student                          & Australia        \\ \hline
P22         & 36           & Trans male      & Mechanical Engineer              & UK               \\ \hline
P23         & 33           & Male            & Software Engineer                & USA              \\ \hline
P24         & 24           & Female          & Student                          & USA              \\ \hline
\end{tabular}
\caption{Demographic details of participants with whom we conducted semi-structured interviews for this study.}
\label{table:participants}
\end{table}

\subsection{Data Collection}
Our interviews lasted an average of 81.65 minutes (sd = 12.16
minutes, range: 60 - 94 minutes) and were conducted over Zoom, recorded, and transcribed. Participants completed an informed consent form before proceeding with the interview.
We explicitly informed participants at the start that we would record the interview to enable analysis and answer our research questions.

To understand the role of personal moderation tools in situated contexts, we first asked participants about their general social media usage during the interview. Next, we asked whether they encountered offensive content on their news feed, what action they took in such cases, and what happened as a result. 
Next, we briefly reviewed the goals of our session and requested a screen-share before asking them to use our website.  We explained that our purpose was not to create a new social network site but to use our simulation of different interfaces as probes to elicit their perceptions of using different personal content moderation tools. We requested that participants ``think-aloud'' \cite{lewis1982using,van1994think} as they tried different interfaces, expressing their likes, dislikes, and confusion and describing additional features they would like to have. 

During these sessions, participants frequently toggled between news feeds and the settings page under each interface to observe how changes to settings altered their feeds. We asked participants to reflect on how they would configure different interfaces if they used them to moderate their news feeds. 
We clarified to the participants that their configurations would affect only the content on their news feed, not other users' feeds. 
In these interviews, we avoided explaining what the different interfaces did without solicitation but explained their purpose and functions when asked. This strategy let us understand the differences between participants' understanding and the actual function of interfaces when such differences existed. 

\begin{figure}
    \centering
    \includegraphics[width=0.65\textwidth]{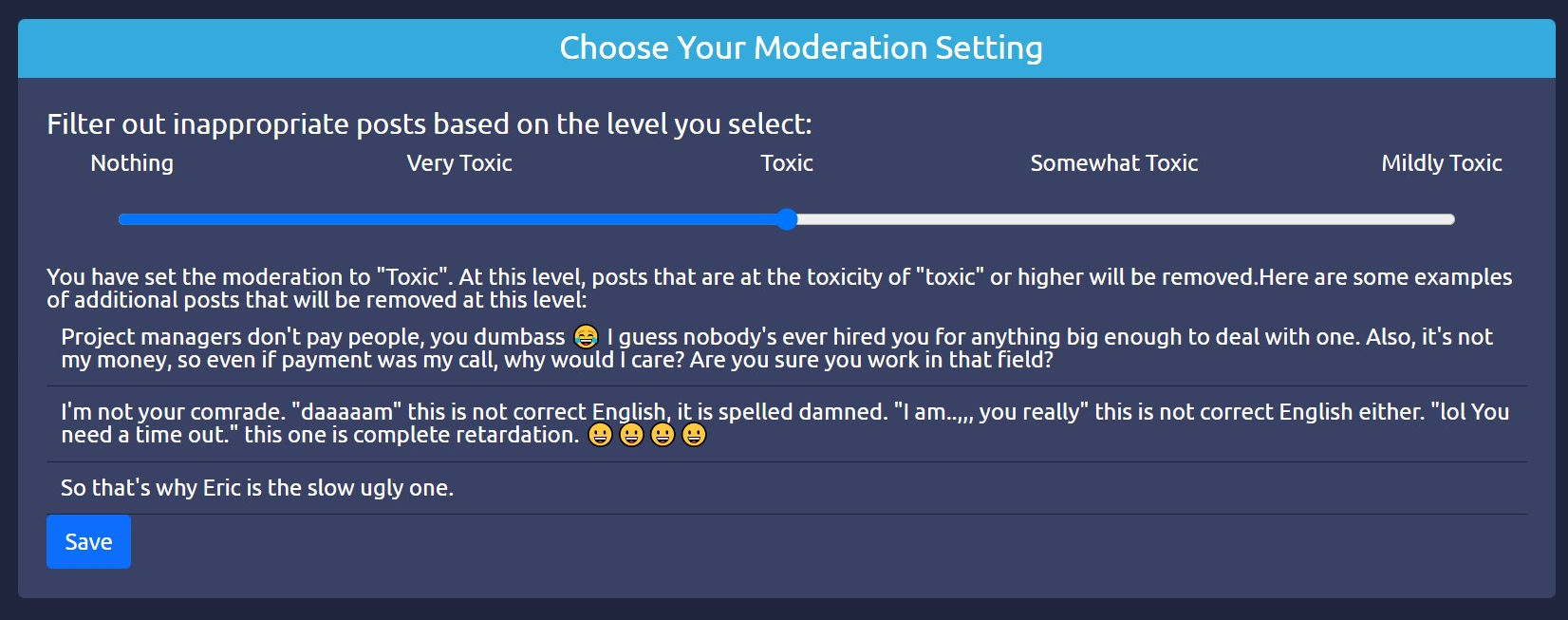}
    \caption{Intensity slider with examples of comments that would be additionally removed when toggling to each level.}
    \label{fig:intensity_eg}
\end{figure}

For our intensity slider, we created an additional interface that, for each level, showed examples of comments that would be removed at that level (Fig. \ref{fig:intensity_eg}). We asked participants to comment on whether these examples influenced their perceptions of this interface.

\begin{figure}
    \centering
    \includegraphics[width=0.35\textwidth]{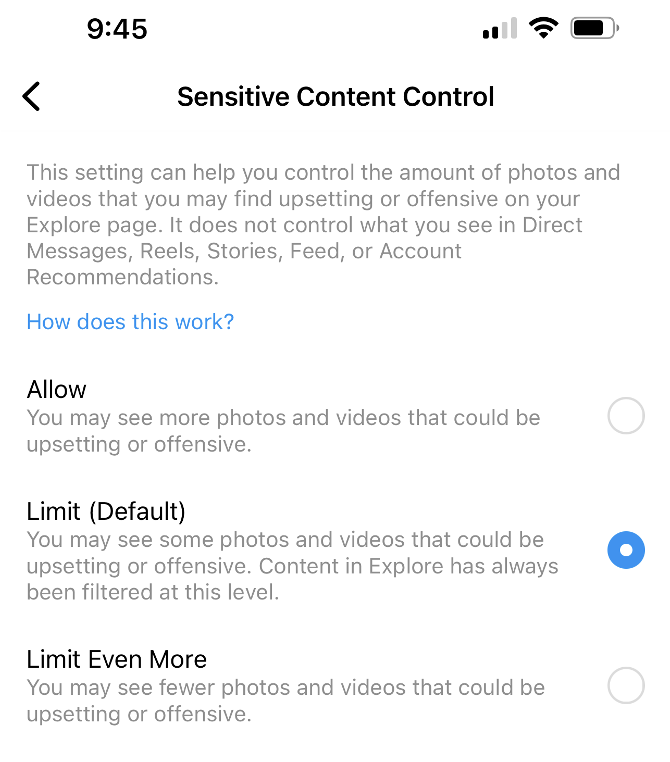}
    \caption{Screenshot of sensitive content control settings implemented by Instagram.}
    \label{fig:instagram}
\end{figure}

In addition to our interface implementations, we showed participants screenshots of categorical toxicity sliders implemented by Intel (Fig. \ref{fig:intel}) and asked them for their thoughts on it. 
We also sought feedback on the sensitive content control settings implemented by Instagram (Fig. \ref{fig:instagram}) since this is a deployed variation of the sliders we implemented on our website.
We encouraged participants to compare and contrast different interfaces based on their efficacy, explainability, and flexibility. 

Immediately following this hands-on session, where participants configured different moderation interfaces, we asked them about their views on the labor involved in setting up such configurations and the role that platforms and policymakers should play in ensuring appropriate content curation. Finally, we asked participants for their demographic information.
After the interview, we compensated each participant with a \$20 Amazon gift card (or an equivalent amount for foreign countries).

\subsection{Analysis}


    


We began our analysis by reading the interview transcripts and familiarizing ourselves with the data. Next, we applied interpretive qualitative analysis to all interview transcripts \cite{merriam2002}. 
We uploaded all transcripts to a dedicated project in Dedoose,\footnote{\url{https://www.dedoose.com}} a cross-platform app for analyzing qualitative research. Then, we conducted ``open coding'' \cite{charmaz2006refs} on a line-by-line basis so that codes stayed close to the data. Through this coding, we categorized segments and created codes that summarized and accounted for each piece of data in concise terms. Examples of codes in this stage included ``Desiring attack notifications'' and ``Seeing offensive content.''



Next, we performed multiple iterative rounds of coding and memo writing. During this process, we continually compared our codes to their associated data. 
To do so, we paid close attention to the dilemmas and tradeoffs that emerged in shaping user preferences, e.g., desiring more control over moderation tools versus exerting more configuration labor.
Our memo writing helped us remain open to emerging themes and deepen our reflections.
The authors discussed the codes, memos, and emerging concepts throughout our analysis each week. 
We contacted some participants to clarify their responses during the analysis stage further.

After the first round of coding, which stayed close to the text, our next round was at a higher level. It resulted in codes such as ``Appreciating the ability to fine-tune'' and ``Labor involved in setting category-based filters.''
Through the following rounds, we combined and distilled our codes into key themes we next present as our findings.

\subsection{Positionality Statement}
As researchers working in the sensitive space of understanding online harm and imagining interventions that may help address it, we briefly reflect on our position on this topic. 
All the authors of this paper feel deeply concerned that online harm is a persistent social problem that disproportionately affects marginalized and vulnerable people \cite{duggan_online_2017}. Some authors also come from marginalized groups and are survivors of harm.
Our prior research on online harassment has helped us understand the inherent limitations of traditional, platform-wide moderation mechanisms; this, in turn, has shaped our desire to look for alternative approaches.
At the same time, we are also concerned with threats to free speech that overly restrictive and globally enacted moderation solutions pose.
In examining and improving the design of personal content moderation tools, we seek to empower users, especially from marginalized groups, to participate in online discourse freely, safely, and on their terms.
\section{Findings}

We begin by describing in Sec. \ref{sec:findings-harms} how participants 
react to experiencing online harms differently, emphasizing the trade-off they face between mitigating harms and missing out on valuable content (RQ 1). Sec. \ref{sec:findings-tools} details the challenges in using personal content moderation tools that participants identified in our interviews (RQ 2).
This subsection offers insights from participants' direct engagement with interview probes, including our toy website and screenshots of existing personal content moderation tools.
Finally, Sec. \ref{sec:findings-labor} and \ref{sec:findings-distribution} present participants' perspectives on the labor involved in moderation configurations and how platforms and lawmakers could better serve them, respectively (RQ 3).

\subsection{Reacting to Online Harms and Concerns about Missing Out on Valuable Content} \label{sec:findings-harms}
Our participants reported a wide range of harms they encountered on social media sites.
They experienced harmful content not just on their news feed, but also in comments to their posts, direct messages, and profile pictures.
They expressed varying responses to such harms and discussed the intention to stay online despite them.
Participants also reflected on the trade-offs between mitigating harmful content and missing out on valuable content. We now detail these reflections.

\subsubsection{Responding to Online Harm}
Participants' responses to experiencing online harm vary depending on their interpretations of online interactions and the tools made available on specific sites. 
These responses included using the available moderation mechanisms, ignoring online harms, and quitting social media.

Our participants showed sufficient literacy of personal moderation affordances available on different platforms. 
Their mental models of what mechanisms such as blocking, muting, and reporting accomplish were aligned with the understanding of authors, who are content moderation experts.
Many participants deployed mechanisms including
reporting or muting offensive posts and reporting, blocking, unfollowing, or muting perpetrator accounts. 
Participants noted that they needed to engage in such actions to protect their mental health. 
For example, P20 began ``hiding'' content he deemed not good for his mental health to avoid repeated exposure. 
However, in line with prior research \cite{kou2021flag,jhaver2018blocklists}, many participants noted unsatisfactory experiences with such actions, especially with reporting inappropriate content.


Some participants preferred not to use any of the moderation mechanisms offered by the platforms.
For example, P7, P21, and P22 preferred to ignore and disengage with harmful content. 
P21 described his reasoning for this behavior as follows:
\begin{quote}
	    \textit{``If one person's being racist, there is probably a lot of other people also doing that, and there's not much of a point for me sifting through every single comment and reporting it. Usually when I see that kind of thing, I just call it [a day] and go do something else.''} – P21
\end{quote}




On the other end, several participants stopped using social media sites altogether due to experiencing harmful content. Both P18 and P22 stopped using Twitter after seeing too much harmful content. P22 used to rely on social media sites to get his news, but due to the negativity on these sites, he turned to view news outside social media. These reactions suggest that efficient, user-friendly personal content moderation tools could benefit some users by letting them address online harms and continue engaging in online discourse.

\subsubsection{Concerns around Personally Moderated Content}
Participants worried about what they might miss if they used any of the interfaces they sampled to censor content for themselves. For example, P24 felt that her social media feeds were already ``too much of an echo chamber'' and  that using content moderation tools would ``over-censor'' what she sees. P14 reflected on his ``fear of missing out'' (FOMO), arguing that he worried more about missing important content than encountering toxic posts. Many participants did not want to shelter themselves on an online platform from the harsh realities of offline life. 
P24 appreciated when Instagram flagged sensitive content,  like images or videos of severely underweight people, but did not stop users from seeing it. 
P11 reported that his social network news feeds once served up a video of a beheading. He was troubled by it but would not want even such gory content to be filtered out:
\begin{quote}
    \textit{``If you go to a party and somebody is talking about that beheading that has happened, I will get exposed to it. It's not like I can cut out those conversations offline.'' - P11}
\end{quote}


Some participants also hesitated to suppress inappropriate posts because they desired to remain informed and take appropriate actions in response to such posts. 
For example, P10, who received rape threats, emphasized the importance of not suppressing threatening posts that could  serve as warning signs for impending offline harm.
P3 noted that if any account she \textit{followed} began posting misogynistic content, she would want to know about it and take actions such as \textit{unfollowing} that account or reconsidering how she related to that person. 
P1 did not mind seeing emotionally volatile posts, even if they were offensive, because she \textit{``wanted to know the reality, like what people are angry about and if there is anything I can do to help them.''}
Three participants felt strongly about their responsibility to keep online spaces safer for everyone by observing and then flagging inappropriate content.
Further, many of our participants were apprehensive about the ability of machine learning-based moderation systems to effectively moderate and therefore preferred the option of minimal or no filtering. 

\subsubsection{Inclination to Remain Open-minded}
Participants also reflected on the balance between avoiding online harms and remaining open-minded. Everyone agreed that egregiously inappropriate content, such as calls to violence or malicious disinformation, went beyond the limits of free speech and should be taken down for everyone.
P11 argued that free speech rights came second to local jurisdiction and that any posts that violated local laws should be removed.
Beyond that, many interviewees felt that platforms should respect the users' First Amendment rights and allow everyone to speak freely. 
As we will discuss further in Section \ref{sec:disc-addressing}, this is a misunderstanding of First Amendment rights since platforms are not arms of the government but private entities that are free to censor anyone they like~\cite{klonick2017,gillespie2015platforms}.

Participants described their desire to remain open-minded when consuming social media content in the context of personal content moderation tools.  
Some argued that personal content moderation controls should not exist because everyone should be more open-minded. 
Overall, participants recognized the difference between their actions to hide a post via personal content moderation vs. a platform's removal of that post for everyone. 
For instance, P14 did not consider his hiding a post as censorship since others could still see that post.

\subsection{Understanding and Engaging with Personal Content Moderation Tools} \label{sec:findings-tools}
Our participants were keen to interact with the various personal content moderation tools we presented to them through our toy website. They also expressed how these tools could address their moderation challenges and improve their social media experience. Many participants were especially excited about the more granular moderation available through Intel's Bleep interface. 
 Since prior research \cite{jhaver_zhang_2022} has documented the need for such tools, we focus here on the nuances of where users see room for improvements in their design.

Our participants reported encountering challenges when engaging with personal content moderation settings. First, they regretted that the terms used to ground various elements of the personal moderation interfaces were often not precisely defined. Second, they felt that the systems' failure to account for critical environmental contexts and offer appropriate levels of granularity hindered their ability to configure aspects of personal content moderation to their liking. Finally, participants reflected on the systems' use of examples to offer explanations and instill transparency. We detail these findings below.

\subsubsection{Defining the Terms and Criteria} \label{sec:findings-define}

We found that participants needed to build a mental model around how the platform carries out personal content moderation before they could more comfortably and deeply engage with moderation settings. While the mental model of how word filters operated---comments containing the configured keywords would be automatically blocked---was generally clear to users, they had more trouble forming mental pictures for the slider and category-based settings. For these settings, platforms present the moderation system as an automated tool that categorizes content into high-level groups such as \textit{``hateful speech''} or \textit{``sensitive content.''} These groups are described by the displayed label and some short text descriptions to direct configuration changes (e.g., ``Feel free to fine tune'' used by Intel, see Fig. \ref{fig:intel}). In such cases, users often desired greater transparency and clarity about the criteria that define these categories. 

\subsubsubsection{Ambiguous Definitions} 
When personal content moderation settings were presented as a categorization of content, either by type (e.g., ``sexism,''  ``racism'') or along a scale (e.g., ``very toxic,'' ``somewhat toxic,'' or ``A little moderation'', ``Some moderation''), participants often wanted to dig deeper into the definitions of these terms. They
found the term \textit{toxicity}, used in our slider-based interfaces, to be especially ambiguous and subjective. For example, P16 commented, \textit{``I can see it working in a black or white manner, but the idea of toxicity---that's very subjective!''}
P7 noted that due to the variability of community norms and topics of interest, \textit{``one thing that can incite hate among one community [can be] kind of benign to another,''} so it can be challenging to understand the platform's criteria. P22 also noted  that different people use language differently, so it may be undesirable if linguistic characteristics, such as the presence of swear words, were used as criteria for toxicity.

Participants also observed this perceived ambiguity of meanings in other interface elements. When examining the Intel category-based filters, both P2 and P7 could not understand how the system defined category groups like ``name-calling'' despite their brief text descriptions in the interface since they found them imprecise. Participants noted that depending on the system's definition and criteria of such terms, they might not consider some instances of name-calling as harmful or offensive, which would guide their configuration actions. However, the lack of clarity in descriptions of relevant terms interfered with their sensemaking.

Besides the confusion around the meanings of terms like ``toxicity'' and ``name-calling,'' participants also reported challenges in understanding what distinctions existed for the levels offered on slider scales (``What is in each level?''). For example, P22 expressed confusion around the classification process for levels on a slider control. It was important for participants to understand what each level meant in more concrete terms than the interface presented to determine whether this type of control was relevant to their moderation goals. Responding to a question about whether there was any additional information that could help him select a level on a slider interface, P22 responded:

\begin{quote}
    \textit{``I mean, kind of, describe how it selects, how it is actually working! [Pointing to a comment] Because thinking about this one---they direct insulted someone. Well, that's pretty rude! [Changing settings and pointing to another comment on the changed news feed] But yeah, that one is calling someone a dumbass as well, so I do not know why this one is seen as worse than that one.''} - P22
\end{quote}  

\subsubsubsection{Evolving Definitions}
Participants also pointed out that meanings of words and criteria used to define categories can evolve, so a static definition (such as \textit{toxicity}) may need to be updated. When discussing the definition of \textit{aggression} in one of the moderation interfaces we showed, P19 commented:
\begin{quote}
    \textit{``You know, something like aggression, you can define aggression on an algorithm. But ableism, you know, [and] LGBTQ discrimination---those are constantly evolving concepts and ideas based on social parameters. And so, having a scale...it will be obsolete the moment it is rolled out. And it will be based on ideology, rather than actual concepts.''} - P19
\end{quote}

\subsubsubsection{Lack of Transparency about Relevant Moderation Criteria}
Participants were curious about the factors that moderation systems incorporate in their decision-making and how they resolve conflicting or competing preferences. 
For example, P5 wanted to know what is included in the input to the algorithms that determine toxicity, asking: ``\textit{How [is] extreme toxicity [determined]? Is this one classified depending on a word, or an expression, or the entire tweet itself? The sentiment of the tweet itself?}''
Three participants assumed toxicity was determined based on the language used and described instances where it is inappropriate to operationalize toxicity based on the presence of certain words. 
For example, P19 felt, \textit{``What makes something toxic isn't so much the word choices, but rather the emotion behind which drives the word choice.''}
Regarding interactions between overlapping criteria, as in the Intel moderation interface, P3 desired more clarity in scenarios where two topical filters may be in effect simultaneously:
\begin{quote}
    ``\textit{I think there's overlap in some of these categories. And so if I don't mind seeing name calling generally, but maybe I don't want to see anything having to do with misogyny, but there's going to be some overlap between those two categories. And then do I not see stuff? Or do I always see stuff? That is confusing to me. How are you to sort out aggression versus name-calling?}'' - P3
\end{quote}

\subsubsection{Failure to Account for Environmental Context}

Participants noted that many personal moderation systems did not or could not account for relevant context when moderating content, yet their desired moderation outcomes were often context-dependent.

\subsubsubsection{Rigidity in Word Filters}
In the preceding sections, we noted that participants found word filters easy to form a mental model around. However, their effects on news feeds were often undesirable in practice due to their inability to account for context. P21 gave an example of how this fluidity of word meanings could be a challenge when moderating news feeds in different cultural communities:
\begin{quote}
    \textit{``I think a lot of these words that have been used depend on the context, like `c*nt.' So, I'm in Australia, we say that all the time and usually it's not a bad thing.''} - P21
\end{quote} 

This lack of ability to account for environmental context can affect not just neighborhood communities but also far-flung communities defined by shared subcultures or social concerns.
\begin{quote}
  \textit{``You can say the exact same sentence but in one context, it is body shaming and in another context it is a compliment for somebody.''} - P15
\end{quote}

\subsubsubsection{Accommodating Use of Reclaimed Slurs}
Concerns about the context also arose for minority communities who reclaimed slurs to fight against historical oppression. For example, some participants speculated that the ``n-word toggle'' from Intel's category-based filters could be problematic for African-American communities that do not object to its use in in-group conversations. While they may wish to apply such a filter to content outside the community, an across-the-board filter that does not account for the poster's identity would also censor well-meaning conversations inside their community. 
Word filters enact a similarly crude moderation, where the only criterion for comment removal is the presence of any configured keyword regardless of other factors.

Insufficient incorporation of relevant context was also cited as a shortcoming of slider-based tools. P2 gave two examples of how inappropriate design of moderation tools can harm rather than help specific communities when commenting on the \textit{body shaming} and \textit{name-calling} categories of the Intel slider: 
\begin{quote}
    \textit{``The fat community is really moving forward with fat empowerment, and you don't want to shut down their posts because they're talking about fatness in positive manner.''} - P2
\end{quote}
\begin{quote}
    \textit{``A straight person can call a gay person a twink and that's like derogatory. But like if a gay person calls another gay person that, that might not be derogatory. One of my best friends is trans and they'll use that term all the time when we're referring to their friends, that's just like how they talk about their friends.''} - P2
\end{quote}

In summary, when user moderation needs depend on environmental context, engaging with personal moderation systems that cannot account for such context or do not allow specifying context as part of the setting can be frustrating.

\subsubsection{Granularity of Control}

Another issue participants noted was that the granularity of control offered by various personal content moderation tools was often inappropriate. This problem was reflected not only in a lack of higher granularity but also through systems offering control over aspects that did not match the users' moderation goals.

\subsubsubsection{Slider-based Controls}
Some participants desired a higher granularity of control when interacting with personal moderation tools. After using a simple toggle to control toxicity moderation, P4 and P17 felt that the moderation seemed too extreme when turned on, leaving the content ``way too overly positive.'' Most participants appreciated the increased degrees of freedom that sliders and category-based controls offered compared to a binary setting. For example, P2 and P11 appreciated the ability to adjust moderation levels across specific categories using Intel filters, noting that these sliders accommodated users with different levels of acceptance for each category. P21 called sliders a good solution ``because you can kind of fine-tune it'' but wished for even greater granularity, suggesting using a continuous, rather than a discrete, slider. 

However, optimal granularity is not achieved simply by increasing the degrees of freedom in all cases. 
Some participants noted that adequate visibility into the behavior of moderation tools was a prerequisite to determining how much control was practicably achievable or even desirable. When differences between settings were not easily interpretable, participants saw little benefit in extra granularity. For example, when asked about the number of levels appropriate for the category slider, one participant noted:
\begin{quote}
    \textit{``When I was looking at the five-level slider, there wasn't a huge difference between 1 and 2, and there wasn't a huge difference between 4 and 5. I would think you might be able to make that just a three level slider and not have a huge difference.''} - P22
\end{quote}

P1, P5, and P6 reported similar concerns for sliders with examples, noting that examples alone often did not provide enough information to infer what was being controlled at the different levels of the sliders.
How we designed our toy platform can possibly partially explain our participants' inference of the lack of differences between adjacent levels.  
However, these responses suggest an essential guideline for any platform to consider when building slider-based moderation tools---they should strike an optimal balance between providing the right granularity and adequate transparency into the differences between various levels created at that granularity. 
P7 commented on the gap between system designers' tendency to simplify moderation settings into levels and the challenge of providing more in-depth, but also complex, control over personal moderation:
\begin{quote}
  ``\textit{It is sort of a web designer's idea of user-friendly in the sense of streamlining everything and putting it on the back-end, but I do not think that is necessarily the same as actually giving users, the client, the levels of choice that they might want, you know? I think, actually, in some ways, a more complex system that doesn't use sort of vague terms like toxicity levels will probably be better in some ways.}'' - P7
\end{quote}



\subsubsubsection{Word Filters}
Issues with granularity also arise with word filters, where including or excluding a single word is a fixed binary choice. 
For example, when word filters are used to moderate swear words, they can present a granularity of control that is too high, exposing users to details that they may not want to face. P9 noted that it can be hard to ``pre-specify'' what type of content he might be uncomfortable with: 
\textit{``You know, you might just not be aware of what you do not want, what's going to bother you.''} Even when participants had specific moderation goals in mind, they lamented that it can be \textit{``a lot of work to think of every possible word that might be offensive''} (P24) and that \textit{``it would be nice to get assistance on covering variants like slight changes in spelling''} (P12). P22 additionally noted that it would not be challenging for bad actors to get around word filters unless the filters had sufficient coverage:
\begin{quote}
    \textit{``I can invent 20, 30 ways to say that word that you just blocked. [People] can write really, really ugly things avoiding any of the words.''} - P22
\end{quote}

Configuring personal moderation at a word-level granularity also forced users to confront harmful content uncomfortably. 
For example, P24 noted that \textit{``I feel, like, very uncomfortable writing...the n-word.''}

\subsubsection{Leveraging Examples to Instill Transparency and Control}

Throughout our interviews, we found the prominent use of example comments to make sense of and control the process of configuring personal moderation. 

\subsubsubsection{Using Examples to Align Mental Models}
Many participants found examples to be a good way to build their mental models about moderation controls, especially when text explanations of relevant terms were missing. For example, P3 and P16 found it challenging to understand the specific meanings of various toxicity levels in intensity filters. However, examples helped them understand the types of comments that each level excludes. P6 additionally expressed that examples not only helped  understand moderation tools but could ease their taking of configuration actions---\textit{``to understand, okay, this is what I have to do!''}

In addition to providing the necessary grounding to contextualize their mental model, participants noted that examples offered an excellent opportunity to identify when their mental model mismatched the actual behaviors of the moderation system. 
%
Some participants reflected on the potential problems with using real examples of offensive or toxic content. For example, P7 noted:
\begin{quote}
    \textit{``It might be a little bit more helpful to show the criteria instead of the examples that would be filtered out because with examples, you have to see the things that you didn't want to see.''}
\end{quote}

\subsubsubsection{Using Examples as a Form of Control}
Examples provided a rich amount of information about the behavior of the moderation tools. As a result, many participants expressed a desire to use examples not just as a passive tool to provide transparency but as a direct means of control. 
For example, P13 foresaw the possibility of using manually written examples to define moderation goals directly or through training models based on labeling existing examples.  
\begin{quote}
    ``\textit{One [way to use examples] is to have me type out an example of something, which I think would be kind of an unpleasant activity and also difficult to imagine right off the top of my head ... The other thing could be, as I scrolled through my timeline, as I encountered pieces of toxic content ... I could put it in these [undesirable] categories.}'' - P13
\end{quote}

\subsubsubsection{Using Examples as Feedback for Control}
In addition to understanding or controlling personal moderation, users relied on real examples to get feedback on their personal moderation settings. In this case, they utilized changes to their feed as examples rather than the examples presented by the system. We observed several participants switching between settings and news feed pages to verify that their configuration of the personal moderation tool achieved their desired goals. 

\begin{quote}
    \textit{``I think, to me, what's really interesting is that this post gets filtered out immediately when you change the moderation setting. I don't understand why it's getting filtered out. So I think... Let me fiddle with it some more.''} - P2
\end{quote}

Indeed, applying and testing out interactive control over moderation became a way for users to form more accurate mental models over personal moderation systems iteratively. 
However, this kind of feedback can be labor intensive:

\begin{quote}
    \textit{``The more effort I invest into something, like the more time I spend, for example, in curating my Twitter feed, the more irritated I get when I get things on my feed that I didn't want.''} - P2
\end{quote}

We discuss these labor aspects in the next section.

\subsection{Labor Involved in Configuring Moderation Systems} \label{sec:findings-labor}

Many participants showed a keen awareness of the extent of labor involved in setting up and maintaining moderation configurations. 
P2 and P3 described this as a trade-off between the extent of agency they have over curating their social media feeds and the labor they must exert to configure moderation filters.
Participants wanted to see improvements to their news feeds in proportion to the effort and time involved in setting up the corresponding configuration. For example, some participants felt frustrated with having the option of multiple levels in the sliders in our study because they kept seeing what they deemed inappropriate comments despite moving between different levels.

A few participants saw setting up moderation configurations as an additional burden they were unwilling to take on. For example, P23 said:
\begin{quote}
    \textit{``I think I am a typical user - I'm probably using my phone at a time I'm really tired, I come back from work---I don't really care, and so if you ask me to customize to that level, my laziness would come out and I'd be like `Nope I don't want to click so much.' ''} - P23
\end{quote}

Similarly, P18 preferred category-based filters over word filters because she used social media to ``decompress'' during her lunch breaks and did not have time to configure the inappropriate keywords worth blocking in word filters.

\begin{quote}
    \textit{``If I have to go on there, and I feel like I'm designing or customizing or engineering and I'm doing this for free…no I don't want to do this, this is supposed to be fun!'' - P18}
\end{quote}

P21 felt that configuring topical category-based filters was onerous. He instead suggested a design with a main toxicity slider, where users could additionally search for and configure specific topic-based filters.
P18 argued that configuring any of the interfaces we showed him would require his full attention; he found these configurations more demanding than other moderation actions, such as flagging a post or blocking a user.
However, he also felt that such configurations would rarely need updating once set up.

Participants also considered the monetary value of labor involved in setting moderation configurations. P2 and P3 pointed out that in enacting moderation configurations, users produce valuable training data for the complex algorithmic systems platforms use and should therefore receive some compensation. 

Some participants wished platforms would design ways to reduce the labor involved in moderation configurations. For example, P3 wanted word filters to have the ability to quickly configure entire groups of pre-configured keywords representing offensive categories such as sexism or homophobia. 
She also wanted to see which keywords her friends added to get ideas about the keywords to configure for her account. 
P1 wanted slider interfaces to indicate how the news feed changes in response to settings changes to reduce the work of going back and forth between the tool and the feed to observe what has changed.





\subsection{Distribution of Responsibility}\label{sec:findings-distribution}
As a follow-up to our participants expressing their views about the different personal moderation tools, we asked them about their perspectives on the obligation of configuring such tools. Specifically, we asked for their opinions on how the responsibility for appropriate content curation should be distributed between the platform and the end users.

In response to this line of inquiry, most participants argued that platforms should, at a minimum, remove the most egregious content posted on the site. 
For example, P12 and P21 noted that many user groups, such as children or technically illiterate users, may be unable to set up their personal moderation preferences. Therefore, platforms are responsible for ensuring appropriate content delivery by default.
P15 and P19 felt that given the secretive nature of content curation that platforms use, they have a social and moral obligation to remove inappropriate comments.
P8 said platforms should not allow social unrest, like the January 6 Capitol attack, to be instigated on their site because it violates their policies. 

\begin{quote}
    \textit{``They should make sure that people stick to how Instagram and other sites originally intended users to share content. But like, if they're using it to incite people, to rile them up purposefully, or to just like using the platform in ways that are harmful, that is going to cause real harm. That becomes the responsibility of the platform. Because these people are like, you know, not using what they built properly.''} - P8
\end{quote}

In the same vein, P2 argued that platforms should consider the ongoing misinformation and hate speech trends and ensure that such content is regulated by default, regardless of users' personal configurations:

\begin{quote}
    \textit{``I think that there are some things like public health misinformation that should just be automatically turned on. Like during COVID, there were a lot of hate crimes towards Asian people. I think platforms, especially large platforms, they know what's going on in the news cycle. They can curate certain moderation categories that are just automatically turned on because they know that misinformation's more likely to be higher in these spaces.''} - P2
\end{quote}

While participants wanted platforms to implement basic site-wide moderation that maintains community integrity and safety, they also wanted access to personal moderation tools to curate their individual experiences. 
For example, P6 and P9 noted that personal moderation tools would empower end users by giving them more control over what they see.
P17 pointed out that the more specificity personal moderation tools allow users to dictate, the more helpful they would be.
P22 considered it in platforms' interest to offer personal moderation tools because such tools let users shape their curation preferences while ensuring that platforms do not suffer any accusations of censorship.

We observed that many participants were willing to spend time and effort configuring their personal preferences for moderation. This readiness is likely because participants recognized that their personal preferences for content curation were niche and could not be captured by site-wide content curation systems. 
%
%
Some participants noted that platforms' site-wide policies and practices were often created in the context of Western culture and might not address the content curation needs of individuals in other cultures. 
This finding is in line with prior research showing that other moderation policies and mechanisms also often fail to account for localized contexts
\cite{pater2016,tushnet2019content,wilson2020hate,jhaver2019automated}. Participants, therefore, considered it vital that personal moderation tools be offered so users can shape their own individualized, culture-aware preferences. 

\begin{quote}
    \textit{``I think it's a combination of both platforms and users [that should decide content curation] because otherwise the thing on the platform typically again is that it's going to be a team of majority white guys in some corner of the world deciding how an Indian boy is going to get his content.''} - P10
\end{quote}

While control over content curation is highly valued, most participants admitted that there should be restrictions on how personal moderation can shape their news feed.
For example, there was a consensus that content that is politically neutral and beneficial for society, such as news of missing children or health information, should always be allowed, regardless of participants' personal moderation preferences.

Interestingly, many participants reported a lack of trust in platforms' motives, a sentiment primarily shaped by reading news reports about how company leaders have handled misinformation and online abuse. 
This loss of trust pushed some participants to seek out and rely on third-party moderation tools, especially those developed by universities or non-profit companies that are more positively perceived. 


Many participants considered that in addition to platforms, end users, and third-party developers, lawmakers also have a role in ensuring appropriate news feeds. P15 thought governments must understand platforms' algorithms to process users' content. P14 and P21 feared that government intervention would increase bureaucracy and politically motivated censorship; they both proposed that a body of lawmakers from different countries be instituted to govern platforms. P21 said:

\begin{quote}
    \textit{``Platforms need to be governed in the same way you would govern a sovereign nation, because they're almost an extension of our society, they're no longer just like a service that we use. I don't really trust the corporate structure to be able to police that system, just because they don't have that sort of incentive.''} - P21
\end{quote}

\section{Discussion}

\subsection{Addressing Harms while Still Enabling Freedom of Speech} \label{sec:disc-addressing}

\subsubsection{Effectively Mitigating Potential Harms}
Our analysis shows that individuals have different preferences regarding encountering offensive speech on social media. This observation is in line with prior research, which shows that while viewing extremist content on social media can be psychologically distressing for most, it is not experienced as harmful by everyone. Some even consider it an awareness-inducing experience that makes them feel better \cite{stubbs2022investigating,tait2008pornographies}.
Since users have varying moderation needs in response to the same content, platforms must prioritize the development of more advanced personal moderation solutions to facilitate individualized experiences and educate users about their utility.

Our participants' desire for such tools supports this call to action. 
As our participant responses show, currently offered moderation tools can be vastly improved by introducing search and analytics features that support easier configuration and selection. 
Prior research on moderation tools for content creators \cite{mallari2021understanding,jhaver2022filterbuddy} and moderators \cite{jhaver2019automated,chandrasekharan2019crossmod} offers blueprints for designing and improving personal content moderation tools.
For example, in line with the templates offered by FilterBuddy \cite{jhaver2022filterbuddy},  category-based filters of the kind featured by Intel could be improved by configuring additional, carefully curated, culture-aware categories in partnership with trusted individuals or experts and allowing users to search for and select them.
Thus, personal content moderation tools could address the challenge of incorporating cultural context in moderation, as highlighted by prior research \cite{pater2016,tushnet2019content,wilson2020hate}.
We found that online harms are not limited to news feeds but can also occur in direct messages and user profiles (Sec. \ref{sec:findings-harms}). 
Therefore, platforms must consider designing personal moderation tools that let users control their experience beyond news feeds.
In fact, our analysis suggests that some users may prefer to use these tools to tune their search results but not their news feeds. 


As our findings show, the fear of missing out (FOMO) makes users wary of over-moderating in response to harm. 
Thus, the challenge is to develop context-aware moderation solutions that understand users' values and safety requirements and then balance the competing needs.
One analysis suggests that besides content removals, another promising approach is deploying content labels, such as fact-checks \cite{zannettou2021won}, interstitial warnings \cite{chandrasekharan2022quarantined}, or content sensitivity alerts, which can serve as reasonable middle-ground solutions to address users' complex needs \cite{morrow2021emerging}.

\subsubsection{Impact on Freedom of Speech}
Personal moderation tools allow end users to control only what they see and do not restrict others' freedom of speech. In a few cases, participants muddied this distinction. 
Indeed, the broader use of these tools could \textit{increase} free speech on platforms, as platforms may be more willing to leave borderline content up if users had an easy way to set personal preferences.
Therefore, adding our voice to previous calls for educating users about moderation policies and related laws \cite{jhaver2019survey,suzor2019we}, we recommend that users be informed about the possibility and bounds of personal moderation tools.

In some cases where participants spoke about ``free speech,'' they objected to \textit{other} users having the option to reduce their view of that content, i.e., they think everybody should be more open-minded.
 %
This discursive production of free speech as an argument against personal moderation tools expresses a moral position.
It dictates the value of not shutting down certain viewpoints, but also, more crucially, requiring everyone else to do the same. 
This perspective suggests that even introducing optional personal moderation tools might engender some resistance.

A separate moral issue that specifically comes up in the case of slider controls is people's concerns with personal moderation controls that lets one see \textit{more} of something offensive or harmful. 
Indeed, much of the outrage expressed online over Intel's Bleep tool focused on the fact that one could say they wanted the maximum level of some objectionable content.
However, we note that this is relative, with default placement potentially having a major impact on perception. 
In addition, framing what the slider accomplishes and the resulting slider labels for levels may also shift user opinion. Such framing may have been why Intel's sliders got significant pushback while a similar category-based 5-level slider feature on Twitch\footnote{\url{https://help.twitch.tv/s/article/how-to-use-automod?language=en_US}} for moderating chat messages in a Twitch stream did not. In the Intel case, one could specify, for instance, that they wanted ``most'' or ``all'' instances of aggression. In contrast, in the Twitch tool, the slider is flipped so that one can specify that they wanted ``more'' or ``maximum'' \textit{filtering out} of aggression.


\subsection{Informing Mental Models of Personal Content Moderation and Improving Controls}

\subsubsection{Clearer Definitions of Harm Categories}
Our analyses show a crucial need for platforms offering personal content moderation tools to clearly and effectively communicate the meanings of their interface elements so that users can form accurate mental models of the type of moderation afforded and the degree of control they have. As we heard from participants, the use of generic, broadly defined keywords, such as `toxicity,' `moderation,' and `name-calling,' raises questions in users' minds about whether their definitions aligned with those of the platform. Based on participant input, we suggest that platforms address this problem by offering more detailed transparency \cite{sinha2002role} into how they define and operationalize the moderation around abstract terms. For instance, if a platform decides to provide controls on moderating \textit{toxicity}, they should offer additional information, like their working definition of toxicity and examples of what they consider toxic. 
In the same vein, when designing sliders that categorize content by type (e.g., ``sexism'', ``racism'') or along a scale (e.g., ``very toxic,'' ``somewhat toxic''), detailed information about what each category includes and how each level differs from others would be valuable.
Further, when social or cultural trends necessitate updates in the system's operationalization of general moderation-related terms like toxicity, they should be communicated to the end users.

Of course, making definitions more grounded can be a fraught process since platforms would open themselves up to criticism from various individuals and communities who may have different expectations for what should be moderated.
As Kumar et al. show, individuals frequently disagree on whether a comment is toxic and which subcategory (e.g., a threat versus an insult) a comment belongs to \cite{kumar2021designing}.
Platforms 
usually hide the massive machinery of content moderation to maintain a veneer of neutrality 
\cite{pasquale2015black}.
However, this secrecy ultimately works against them as it leads to suspicions that platforms are biased \cite{myers2018censored}.
We argue that it is better to provide an imperfect but transparent moderation tool and hash out policy debates in the public  than to provide a system that is opaque and unusable only to avoid controversy.

\subsubsection{Transparency around the Algorithms Behind Controls}
We found that in addition to definitions, users also desire more information about the inner logic and inputs to algorithms behind personal moderation tools. However, platforms usually keep this information opaque by claiming a need to protect their intellectual property  \cite{gillespie2018custodians,suzor2019we}. 
This information is necessary for users to avoid violating their expectations, which may contribute to a loss of trust \cite{kizilcec2016much}.
Prior research has argued that instead of seamless interfaces, designing certain seams into an algorithmic system can affect users' understanding of that system and their interactions with it \cite{chalmers2003seamful,eslami2016first,jhaver2018airbnb}.
Thus, designers can experiment with building carefully designed
seams, such as the main determining factors for moderation decisions, that expose more about algorithms behind personal moderation tools and examine users' responses.

While our findings show that offering visibility into moderation processes is crucial, this may require an overhaul of how personal moderation tools currently work behind the scenes. 
The internal logic of algorithms driving these tools is unknown, but their primary goal likely is to maximize toxicity classification accuracy, and they do not prioritize explainability.
In the transition to making systems more interpretable, personal moderation tools would inevitably need to be more accountable.
Prior research on related moderation mechanisms such as post-removals, flagging systems, and appeal procedures against moderation decisions have also called for platforms to be more accountable \cite{crawford2016,kou2021flag,myers2018censored,vaccaro2021cont,vaccaro2020at}. 
We add to these calls and offer specific recommendations for personal moderation tools. We suggest that designers alter the underlying algorithms at work so that their outputs are more understandable to the end user and in line with the definitional information provided by them to the greatest possible extent.

We found that participants often hypothesized about how personal moderation systems operated, forming their own (often inaccurate) folk theories \cite{devito2018people,jhaver2018airbnb} about our simulated systems.
Our analysis shows that inaccurate folk theories can lead to confusion and potential disappointment with personal moderation tools. 
While the full details of their inner logic may be too complex to be helpful to lay users, these tools do not even offer any \textit{justification} \cite{tintarev2011designing,vig2009tagsplanations} for their moderation decisions by surfacing the key determining factors. 
Although the field of explainable models and AI is still emerging~\cite{arrieta2020explainable,adadi2018peeking}, more can be done to provide details about the conventional aspects of systems. For example, platforms could clarify how moderation tools' overlapping settings interact, disclose whether systems are primarily rule-based or model-based, and reveal the criteria used during the training and evaluation of any models involved. 
Based on our participants' input, we suggest that using example comments at different stages is valuable to instill greater understanding.
Providing additional transparency can help users decide which aspects of the systems they can trust and rely on and which parts need more caution.

\subsubsection{Providing Desired Granularity and Continuous Adjustment}
Furthermore, we found that while participants desired more control over personal moderation, platforms often failed to provide adequate controls to achieve that. For one, providing more degrees of freedom (like more levels of moderation) does not guarantee improved control over the system. Controls must be backed by a sufficiently crisp mental model of what the moderation system is meant to do and how well it can achieve that~\cite{jannach2019explanations,jannach2016user}. Indeed, we found that the level of clarity in participants' mental models greatly affected how they engaged with personal moderation controls, with some noting that the number of slider levels should be \textit{reduced} because they did not perceive a difference between some levels. Additionally, we noticed a strong desire for moderation tools to account for more context \cite{adomavicius2011context}, like differences between communities and different linguistic patterns of groups and individuals. Designing controls to enable context-dependent moderation is a promising direction for future work. 
For example, interactive moderation tools that let users specify the relevant context (e.g., never hide racist comments from a specific user or group) would increase the utility of such tools.

Finally, we propose that configuring personal moderation be continuous and iterative. For one, the currently limited controllability of static personal moderation tools does not let users fully specify their moderation preferences.
Additionally, some participants indicated that they may not be aware of their moderation needs until they encounter undesirable content. Prior research~\cite{jhaver2022filterbuddy} has shown that users' moderation goals may also evolve over time.
Therefore, platforms should offer mechanisms to adjust moderation in the context of results dynamically \cite{jannach2016user}.
For instance, platforms should look towards providing just-in-time~\cite{golhar1991just} controls, e.g., automatically incorporating the signal that ``reported'' content is undesirable so that users can customize personal moderation when it is most relevant to them.
Our participants' enthusiasm for using examples as a form of control attests to the utility of this approach.
Platforms should also provide feedback mechanisms, so users can systematically evaluate the efficacy of their current settings as opposed to the current ad-hoc reflections on their feed mentioned by our participants.

\subsection{Managing Labor and Distribution of Responsibility}
Our findings suggest that users keenly attend to the extent of labor required for personal moderation. Excessive cognitive demands of moderation configurations can deter users from engaging with personal moderation tools. Therefore, designers must devise \textit{efficient} solutions that let users quickly configure their moderation preferences while retaining granular control. One recent example of such a solution is FilterBuddy, a word filter tool for YouTube that lets users quickly import entire pre-built categories of offensive keywords, e.g., about sexism and racism, but allow subsequent configuration changes for each phrase \cite{jhaver2022filterbuddy}. 
Given the tension between users' needs to minimize labor and improve control, designing tools that let users configure their preferences at varying levels of specificity offers a promising direction.

As our findings show, relying on like-minded others for co-creating shared moderation preferences can also reduce moderation labor.
Further, showing improvement metrics in news feeds, e.g., the number of offensive comments hidden, can also encourage users to engage with moderation tools.
Once appropriately configured, many personal moderation tools may need only occasional tweaks.
Thus, platforms can emphasize the utility of a one-time investment in configuring moderation tools and make it easier for newcomers to understand and engage with them.

Regardless of how much platforms promote engagement with personal moderation, our analysis predicts that some users would hesitate to invest any time in changing their moderation settings, depicting a  tendency to use the default settings \cite{dhingra2012default,shah2008software}, while others may be unable to do so because of their lack of digital literacy. In addition to improving personal moderation capabilities, it is therefore vital that platforms provide sensible defaults.
This recommendation is buttressed by our finding that most users expect a baseline level of platform review that would catch and remove blatantly inappropriate posts. However, platforms often fail to meet this expectation:  our participants' frequent experiences of online harms speak to the urgency and importance of this effort.
Thus, personal moderation tools do \textit{not} absolve platforms of the necessity to conduct moderation. We argue against proposals that put all the onus on end users to personally filter content for themselves, as this may exacerbate the digital divide between those who can and cannot configure personal moderation settings.

In addition to platforms, third-party developers can also improve online spaces by implementing innovative moderation solutions. Our findings suggest that users are likelier to use tools built by academics and reliable non-profits. Moreover, policymakers can incentivize platforms to invest in creating new personal moderation interfaces or supporting the community of third-party developers. 
For this to occur, lawmakers require a better appreciation of what is at stake and a better understanding of how platforms implement personal moderation tools.

\subsection{Limitations and Future Work}

We focus on removing ``toxicity'' in three of our controls in our simulated social media feed. 
We chose this due to the availability of social media datasets with human-annotated labels for toxicity and the prevalence of the ``toxicity'' concept in academia and industry. 
However, personal content moderation can also include other types of content that may be undesirable to some, such as spam, sexually suggestive or explicit content, or violent content. 
It would be interesting to explore differing user preferences regarding these other categories of content and to understand the kinds of categories that users would like to adjust separately.

We deployed a toy web application that shows only 30 comments at a time, and it does not incorporate the effects of personalized content streams that users generally have within their social media feeds. Therefore, longitudinally studying user interactions with tools on real platforms hosting many more comments should reveal additional insights.
The simulated controls we built analyze only text, and our feed consists only of text comments. 
These controls could moderate other content, including audio, image, and video. The Bleep tool, as an example, focuses on filtering out audio in audio-based gaming chat rooms. Future studies could specifically examine personal moderation of audio, image, and video content, which may have different user preferences and considerations. For instance, violence and gore in visual content might be a higher priority for filtering since they may be more viscerally harmful.
Our design exploration and toy website creation did not focus on a specific social media site but sought inspiration from various platforms. Since these tools are available alongside the platform-offered moderation mechanisms on each site, platform-specific explorations are needed to examine further the situated use and perspectives on personal moderation tools. 


The specific terms (e.g., toxicity) that our toy platform uses could have influenced our findings about ambiguous definitions (Sec. \ref{sec:findings-define}). However, we note that Twitch and Instagram also use similarly abstract terms on their moderation sliders, such as `offensive' and `aggression,' that are likely to engender ambiguity. Future work that tests users on specific platforms would clarify the extent to which this problem persists in currently deployed tools. 

Our method consisted of conducting semi-structured with 24 participants. While this let us elicit a wide range of concerns, user preferences, and nuances in engagement with personal moderation tools,
the small sample size limited us from asking questions about the relative popularity of the different interfaces we tested. 
Going forward, we plan to conduct large-scale surveys with samples representative of the general population of internet users to answer such questions.
\section{Conclusion}
As a one-size-fits-all model for content moderation is insufficient, platforms need to consider the tools they provide end-users, so they can customize their moderation beyond what is caught at the platform level.
We call this \textit{personal moderation} and identify its two variations: \textit{personal account moderation} and \textit{personal content moderation}. We examine users' preferences regarding personal content moderation tools in this research.
Our analysis shows that these tools would benefit from providing greater context awareness, clarity in the meanings of their interface elements, and justifications behind their decisions. 
Offering these tools does not exempt platforms from ensuring the efficacy of their baseline moderation. 
However, these tools can let users customize their social media experience without infringing
on free speech concerns. 
Policymakers should also compel platforms to invest in building and supporting innovative personal content moderation tools.

\begin{acks}
We thank the participants who took part in our interviews. In addition, we thank the members of the Social Futures Lab and the DUB Summer REU program for giving feedback on this project, as well as Sandy Kaplan for her help with editing and proofreading the paper. Alice Zhang was supported during this work as part of the CRA DREU internship program. This work was also partly supported by a Google Research Scholar award.
\end{acks}

\bibliographystyle{ACM-Reference-Format}
\bibliography{00-sample-manuscript}


\begin{thebibliography}{119}


\ifx \showCODEN    \undefined \def \showCODEN     #1{\unskip}     \fi
\ifx \showDOI      \undefined \def \showDOI       #1{#1}\fi
\ifx \showISBNx    \undefined \def \showISBNx     #1{\unskip}     \fi
\ifx \showISBNxiii \undefined \def \showISBNxiii  #1{\unskip}     \fi
\ifx \showISSN     \undefined \def \showISSN      #1{\unskip}     \fi
\ifx \showLCCN     \undefined \def \showLCCN      #1{\unskip}     \fi
\ifx \shownote     \undefined \def \shownote      #1{#1}          \fi
\ifx \showarticletitle \undefined \def \showarticletitle #1{#1}   \fi
\ifx \showURL      \undefined \def \showURL       {\relax}        \fi
\providecommand\bibfield[2]{#2}
\providecommand\bibinfo[2]{#2}
\providecommand\natexlab[1]{#1}
\providecommand\showeprint[2][]{arXiv:#2}

\bibitem[Abdul-Rahman and Hailes(2000)]%
        {abdul2000supporting}
\bibfield{author}{\bibinfo{person}{Alfarez Abdul-Rahman} {and}
  \bibinfo{person}{Stephen Hailes}.} \bibinfo{year}{2000}\natexlab{}.
\newblock \showarticletitle{Supporting trust in virtual communities}. In
  \bibinfo{booktitle}{\emph{Proceedings of the 33rd annual Hawaii international
  conference on system sciences}}. IEEE, \bibinfo{pages}{9--pp}.
\newblock


\bibitem[Adadi and Berrada(2018)]%
        {adadi2018peeking}
\bibfield{author}{\bibinfo{person}{Amina Adadi} {and} \bibinfo{person}{Mohammed
  Berrada}.} \bibinfo{year}{2018}\natexlab{}.
\newblock \showarticletitle{Peeking inside the black-box: a survey on
  explainable artificial intelligence (XAI)}.
\newblock \bibinfo{journal}{\emph{IEEE access}}  \bibinfo{volume}{6}
  (\bibinfo{year}{2018}), \bibinfo{pages}{52138--52160}.
\newblock


\bibitem[Adomavicius and Tuzhilin(2005)]%
        {adomavicius2005toward}
\bibfield{author}{\bibinfo{person}{Gediminas Adomavicius} {and}
  \bibinfo{person}{Alexander Tuzhilin}.} \bibinfo{year}{2005}\natexlab{}.
\newblock \showarticletitle{Toward the next generation of recommender systems:
  A survey of the state-of-the-art and possible extensions}.
\newblock \bibinfo{journal}{\emph{IEEE transactions on knowledge and data
  engineering}} \bibinfo{volume}{17}, \bibinfo{number}{6}
  (\bibinfo{year}{2005}), \bibinfo{pages}{734--749}.
\newblock


\bibitem[Adomavicius and Tuzhilin(2011)]%
        {adomavicius2011context}
\bibfield{author}{\bibinfo{person}{Gediminas Adomavicius} {and}
  \bibinfo{person}{Alexander Tuzhilin}.} \bibinfo{year}{2011}\natexlab{}.
\newblock \showarticletitle{Context-aware recommender systems}.
\newblock In \bibinfo{booktitle}{\emph{Recommender systems handbook}}.
  \bibinfo{publisher}{Springer}, \bibinfo{pages}{217--253}.
\newblock


\bibitem[Andjelkovic et~al\mbox{.}(2016)]%
        {andjelkovic2016moodplay}
\bibfield{author}{\bibinfo{person}{Ivana Andjelkovic}, \bibinfo{person}{Denis
  Parra}, {and} \bibinfo{person}{John O'Donovan}.}
  \bibinfo{year}{2016}\natexlab{}.
\newblock \showarticletitle{Moodplay: Interactive mood-based music discovery
  and recommendation}. In \bibinfo{booktitle}{\emph{Proceedings of the 2016
  conference on user modeling adaptation and personalization}}.
  \bibinfo{pages}{275--279}.
\newblock


\bibitem[Arrieta et~al\mbox{.}(2020)]%
        {arrieta2020explainable}
\bibfield{author}{\bibinfo{person}{Alejandro~Barredo Arrieta},
  \bibinfo{person}{Natalia D{\'\i}az-Rodr{\'\i}guez}, \bibinfo{person}{Javier
  Del~Ser}, \bibinfo{person}{Adrien Bennetot}, \bibinfo{person}{Siham Tabik},
  \bibinfo{person}{Alberto Barbado}, \bibinfo{person}{Salvador Garc{\'\i}a},
  \bibinfo{person}{Sergio Gil-L{\'o}pez}, \bibinfo{person}{Daniel Molina},
  \bibinfo{person}{Richard Benjamins}, {et~al\mbox{.}}}
  \bibinfo{year}{2020}\natexlab{}.
\newblock \showarticletitle{Explainable Artificial Intelligence (XAI):
  Concepts, taxonomies, opportunities and challenges toward responsible AI}.
\newblock \bibinfo{journal}{\emph{Information fusion}}  \bibinfo{volume}{58}
  (\bibinfo{year}{2020}), \bibinfo{pages}{82--115}.
\newblock


\bibitem[Bahrini and Qaffas(2019)]%
        {bahrini2019impact}
\bibfield{author}{\bibinfo{person}{Ra{\'e}f Bahrini} {and}
  \bibinfo{person}{Alaa~A Qaffas}.} \bibinfo{year}{2019}\natexlab{}.
\newblock \showarticletitle{Impact of information and communication technology
  on economic growth: Evidence from developing countries}.
\newblock \bibinfo{journal}{\emph{Economies}} \bibinfo{volume}{7},
  \bibinfo{number}{1} (\bibinfo{year}{2019}), \bibinfo{pages}{21}.
\newblock


\bibitem[Balkin(2013)]%
        {balkin2013old}
\bibfield{author}{\bibinfo{person}{Jack~M Balkin}.}
  \bibinfo{year}{2013}\natexlab{}.
\newblock \showarticletitle{Old-school/new-school speech regulation}.
\newblock \bibinfo{journal}{\emph{Harv. L. Rev.}}  \bibinfo{volume}{127}
  (\bibinfo{year}{2013}), \bibinfo{pages}{2296}.
\newblock


\bibitem[Bastien(2021)]%
        {bastien_2021}
\bibfield{author}{\bibinfo{person}{Bastien}.} \bibinfo{year}{2021}\natexlab{}.
\newblock \bibinfo{title}{Customized content moderation: One size doesn't fit
  all}.
\newblock
\newblock
\urldef\tempurl%
\url{https://www.bodyguard.ai/blog/customized-content-moderation-one-size-doesnt-fit-all}
\showURL{%
\tempurl}


\bibitem[Bhargava et~al\mbox{.}(2019)]%
        {bhargava2019gobo}
\bibfield{author}{\bibinfo{person}{Rahul Bhargava}, \bibinfo{person}{Anna
  Chung}, \bibinfo{person}{Neil~S Gaikwad}, \bibinfo{person}{Alexis Hope},
  \bibinfo{person}{Dennis Jen}, \bibinfo{person}{Jasmin Rubinovitz},
  \bibinfo{person}{Bel{\'e}n Sald{\'\i}as-Fuentes}, {and}
  \bibinfo{person}{Ethan Zuckerman}.} \bibinfo{year}{2019}\natexlab{}.
\newblock \showarticletitle{Gobo: A system for exploring user control of
  invisible algorithms in social media}. In
  \bibinfo{booktitle}{\emph{Conference Companion Publication of the 2019 on
  Computer Supported Cooperative Work and Social Computing}}.
  \bibinfo{pages}{151--155}.
\newblock


\bibitem[Burrell et~al\mbox{.}(2019)]%
        {burrell2019users}
\bibfield{author}{\bibinfo{person}{Jenna Burrell}, \bibinfo{person}{Zoe Kahn},
  \bibinfo{person}{Anne Jonas}, {and} \bibinfo{person}{Daniel Griffin}.}
  \bibinfo{year}{2019}\natexlab{}.
\newblock \showarticletitle{When users control the algorithms: values expressed
  in practices on twitter}.
\newblock \bibinfo{journal}{\emph{Proceedings of the ACM on Human-Computer
  Interaction}} \bibinfo{volume}{3}, \bibinfo{number}{CSCW}
  (\bibinfo{year}{2019}), \bibinfo{pages}{1--20}.
\newblock


\bibitem[Chalmers and MacColl(2003)]%
        {chalmers2003seamful}
\bibfield{author}{\bibinfo{person}{Matthew Chalmers} {and} \bibinfo{person}{Ian
  MacColl}.} \bibinfo{year}{2003}\natexlab{}.
\newblock \showarticletitle{Seamful and seamless design in ubiquitous
  computing}. In \bibinfo{booktitle}{\emph{Workshop at the crossroads: The
  interaction of HCI and systems issues in UbiComp}}, Vol.~\bibinfo{volume}{8}.
\newblock


\bibitem[Chandrasekharan et~al\mbox{.}(2019)]%
        {chandrasekharan2019crossmod}
\bibfield{author}{\bibinfo{person}{Eshwar Chandrasekharan},
  \bibinfo{person}{Chaitrali Gandhi}, \bibinfo{person}{Matthew~Wortley
  Mustelier}, {and} \bibinfo{person}{Eric Gilbert}.}
  \bibinfo{year}{2019}\natexlab{}.
\newblock \showarticletitle{Crossmod: A Cross-Community Learning-based System
  to Assist Reddit Moderators}.
\newblock \bibinfo{journal}{\emph{Proceedings of the ACM on Human-Computer
  Interaction}} \bibinfo{volume}{3}, \bibinfo{number}{CSCW}
  (\bibinfo{year}{2019}), \bibinfo{pages}{1--30}.
\newblock


\bibitem[Chandrasekharan et~al\mbox{.}(2022)]%
        {chandrasekharan2022quarantined}
\bibfield{author}{\bibinfo{person}{Eshwar Chandrasekharan},
  \bibinfo{person}{Shagun Jhaver}, \bibinfo{person}{Amy Bruckman}, {and}
  \bibinfo{person}{Eric Gilbert}.} \bibinfo{year}{2022}\natexlab{}.
\newblock \showarticletitle{Quarantined! Examining the Effects of a
  Community-Wide Moderation Intervention on Reddit}.
\newblock \bibinfo{journal}{\emph{ACM Trans. Comput.-Hum. Interact.}}
  (\bibinfo{year}{2022}).
\newblock
\urldef\tempurl%
\url{https://doi.org/10.1145/3490499}
\showDOI{\tempurl}


\bibitem[Charmaz(2006)]%
        {charmaz2006refs}
\bibfield{author}{\bibinfo{person}{Kathy Charmaz}.}
  \bibinfo{year}{2006}\natexlab{}.
\newblock \bibinfo{booktitle}{\emph{{Constructing grounded theory: a practical
  guide through qualitative analysis}}}.
\newblock
\showISBNx{9780761973522}
\showISSN{07408188}
\urldef\tempurl%
\url{https://doi.org/10.1016/j.lisr.2007.11.003}
\showDOI{\tempurl}
\showeprint[arXiv]{arXiv:1011.1669v3}


\bibitem[Cole(2018)]%
        {vice2}
\bibfield{author}{\bibinfo{person}{Samantha Cole}.}
  \bibinfo{year}{2018}\natexlab{}.
\newblock \bibinfo{title}{Where Did the Concept of 'Shadow Banning' Come From?}
\newblock
\newblock
\urldef\tempurl%
\url{https://www.vice.com/en/article/a3q744/where-did-shadow-banning-come-from-trump-republicans-shadowbanned}
\showURL{%
\tempurl}


\bibitem[Crawford and Gillespie(2016)]%
        {crawford2016}
\bibfield{author}{\bibinfo{person}{Kate Crawford} {and}
  \bibinfo{person}{Tarleton Gillespie}.} \bibinfo{year}{2016}\natexlab{}.
\newblock \showarticletitle{What is a flag for? Social media reporting tools
  and the vocabulary of complaint}.
\newblock \bibinfo{journal}{\emph{New Media \& Society}} \bibinfo{volume}{18},
  \bibinfo{number}{3} (\bibinfo{year}{2016}), \bibinfo{pages}{410--428}.
\newblock
\urldef\tempurl%
\url{https://doi.org/10.1177/1461444814543163}
\showDOI{\tempurl}
\showeprint{https://doi.org/10.1177/1461444814543163}


\bibitem[DeNardis and Hackl(2015)]%
        {denardis2015internet}
\bibfield{author}{\bibinfo{person}{Laura DeNardis} {and}
  \bibinfo{person}{Andrea~M Hackl}.} \bibinfo{year}{2015}\natexlab{}.
\newblock \showarticletitle{Internet governance by social media platforms}.
\newblock \bibinfo{journal}{\emph{Telecommunications Policy}}
  \bibinfo{volume}{39}, \bibinfo{number}{9} (\bibinfo{year}{2015}),
  \bibinfo{pages}{761--770}.
\newblock


\bibitem[DeVito et~al\mbox{.}(2018)]%
        {devito2018people}
\bibfield{author}{\bibinfo{person}{Michael~A DeVito}, \bibinfo{person}{Jeremy
  Birnholtz}, \bibinfo{person}{Jeffery~T Hancock}, \bibinfo{person}{Megan
  French}, {and} \bibinfo{person}{Sunny Liu}.} \bibinfo{year}{2018}\natexlab{}.
\newblock \showarticletitle{How people form folk theories of social media feeds
  and what it means for how we study self-presentation}. In
  \bibinfo{booktitle}{\emph{Proceedings of the 2018 CHI Conference on Human
  Factors in Computing Systems}}. ACM, \bibinfo{pages}{120}.
\newblock


\bibitem[Dhingra et~al\mbox{.}(2012)]%
        {dhingra2012default}
\bibfield{author}{\bibinfo{person}{Nikhil Dhingra}, \bibinfo{person}{Zach
  Gorn}, \bibinfo{person}{Andrew Kener}, {and} \bibinfo{person}{Jason Dana}.}
  \bibinfo{year}{2012}\natexlab{}.
\newblock \showarticletitle{The default pull: An experimental demonstration of
  subtle default effects on preferences}.
\newblock \bibinfo{journal}{\emph{Judgment and Decision Making}}
  \bibinfo{volume}{7}, \bibinfo{number}{1} (\bibinfo{year}{2012}),
  \bibinfo{pages}{69}.
\newblock


\bibitem[Diaz(2021)]%
        {polygon}
\bibfield{author}{\bibinfo{person}{Ana Diaz}.} \bibinfo{year}{2021}\natexlab{}.
\newblock \bibinfo{title}{Intel responds to hate speech tool getting roasted by
  the internet}.
\newblock
\newblock
\urldef\tempurl%
\url{https://www.polygon.com/22374120/intel-bleep-voice-chat-hate-speech-censor-spirit-ai}
\showURL{%
\tempurl}


\bibitem[Dooms et~al\mbox{.}(2014)]%
        {dooms2014improving}
\bibfield{author}{\bibinfo{person}{Simon Dooms}, \bibinfo{person}{Toon
  De~Pessemier}, {and} \bibinfo{person}{Luc Martens}.}
  \bibinfo{year}{2014}\natexlab{}.
\newblock \showarticletitle{Improving IMDb movie recommendations with
  interactive settings and filters}. In \bibinfo{booktitle}{\emph{8th ACM
  Conference on Recommender Systems (Poster-RecSys 2014)}},
  Vol.~\bibinfo{volume}{1247}.
\newblock


\bibitem[Dorsey(2022)]%
        {jackdorsey}
\bibfield{author}{\bibinfo{person}{Jack Dorsey}.}
  \bibinfo{year}{2022}\natexlab{}.
\newblock \bibinfo{title}{a native internet protocol for social media}.
\newblock
\newblock
\urldef\tempurl%
\url{https://www.getrevue.co/profile/jackjack/issues/a-native-internet-protocol-for-social-media-1503112}
\showURL{%
\tempurl}


\bibitem[Dosono and Semaan(2019)]%
        {dosono2019mod}
\bibfield{author}{\bibinfo{person}{Bryan Dosono} {and} \bibinfo{person}{Bryan
  Semaan}.} \bibinfo{year}{2019}\natexlab{}.
\newblock \showarticletitle{Moderation Practices as Emotional Labor in
  Sustaining Online Communities: The Case of AAPI Identity Work on Reddit}. In
  \bibinfo{booktitle}{\emph{Proceedings of the 2019 CHI Conference on Human
  Factors in Computing Systems}} (Glasgow, Scotland Uk)
  \emph{(\bibinfo{series}{CHI '19})}. \bibinfo{publisher}{Association for
  Computing Machinery}, \bibinfo{address}{New York, NY, USA},
  \bibinfo{pages}{1–13}.
\newblock
\showISBNx{9781450359702}
\urldef\tempurl%
\url{https://doi.org/10.1145/3290605.3300372}
\showDOI{\tempurl}


\bibitem[Duggan(2017)]%
        {duggan_online_2017}
\bibfield{author}{\bibinfo{person}{Maeve Duggan}.}
  \bibinfo{year}{2017}\natexlab{}.
\newblock \bibinfo{title}{Online {Harassment} 2017}.
\newblock
\newblock
\urldef\tempurl%
\url{https://www.pewresearch.org/internet/2017/07/11/online-harassment-2017/}
\showURL{%
\tempurl}


\bibitem[Engler(2022)]%
        {engler_2022}
\bibfield{author}{\bibinfo{person}{Maggie Engler}.}
  \bibinfo{year}{2022}\natexlab{}.
\newblock \bibinfo{title}{Middleware and the customization of content
  moderation}.
\newblock
\newblock
\urldef\tempurl%
\url{https://integrityinstitute.org/our-ideas/hear-from-our-fellows/middleware-and-the-customization}
\showURL{%
\tempurl}


\bibitem[Eslami et~al\mbox{.}(2016)]%
        {eslami2016first}
\bibfield{author}{\bibinfo{person}{Motahhare Eslami}, \bibinfo{person}{Karrie
  Karahalios}, \bibinfo{person}{Christian Sandvig}, \bibinfo{person}{Kristen
  Vaccaro}, \bibinfo{person}{Aimee Rickman}, \bibinfo{person}{Kevin Hamilton},
  {and} \bibinfo{person}{Alex Kirlik}.} \bibinfo{year}{2016}\natexlab{}.
\newblock \showarticletitle{First i like it, then i hide it: Folk theories of
  social feeds}. In \bibinfo{booktitle}{\emph{Proceedings of the 2016 cHI
  conference on human factors in computing systems}}. ACM,
  \bibinfo{pages}{2371--2382}.
\newblock


\bibitem[Eslami et~al\mbox{.}(2015)]%
        {eslami2015always}
\bibfield{author}{\bibinfo{person}{Motahhare Eslami}, \bibinfo{person}{Aimee
  Rickman}, \bibinfo{person}{Kristen Vaccaro}, \bibinfo{person}{Amirhossein
  Aleyasen}, \bibinfo{person}{Andy Vuong}, \bibinfo{person}{Karrie Karahalios},
  \bibinfo{person}{Kevin Hamilton}, {and} \bibinfo{person}{Christian Sandvig}.}
  \bibinfo{year}{2015}\natexlab{}.
\newblock \showarticletitle{I always assumed that I wasn't really that close to
  [her]: Reasoning about Invisible Algorithms in News Feeds}. In
  \bibinfo{booktitle}{\emph{Proceedings of the 33rd annual ACM conference on
  human factors in computing systems}}. ACM, \bibinfo{pages}{153--162}.
\newblock


\bibitem[Gault(2021)]%
        {vice}
\bibfield{author}{\bibinfo{person}{Matthew Gault}.}
  \bibinfo{year}{2021}\natexlab{}.
\newblock \bibinfo{title}{Intel's Dystopian Anti-Harassment AI Lets Users Opt
  In for `Some' Racism}.
\newblock
\newblock
\urldef\tempurl%
\url{https://www.vice.com/en/article/dyvgvk/intels-dystopian-anti-harassment-ai-lets-users-opt-in-for-some-racism}
\showURL{%
\tempurl}


\bibitem[Geiger(2016)]%
        {geiger2016}
\bibfield{author}{\bibinfo{person}{R.~Stuart Geiger}.}
  \bibinfo{year}{2016}\natexlab{}.
\newblock \showarticletitle{Bot-based collective blocklists in Twitter: the
  counterpublic moderation of harassment in a networked public space}.
\newblock \bibinfo{journal}{\emph{Information, Communication \& Society}}
  \bibinfo{volume}{19}, \bibinfo{number}{6} (\bibinfo{year}{2016}),
  \bibinfo{pages}{787--803}.
\newblock
\urldef\tempurl%
\url{https://doi.org/10.1080/1369118X.2016.1153700}
\showDOI{\tempurl}
\showeprint{https://doi.org/10.1080/1369118X.2016.1153700}


\bibitem[Gillespie(2015)]%
        {gillespie2015platforms}
\bibfield{author}{\bibinfo{person}{Tarleton Gillespie}.}
  \bibinfo{year}{2015}\natexlab{}.
\newblock \showarticletitle{Platforms intervene}.
\newblock \bibinfo{journal}{\emph{Social Media+ Society}} \bibinfo{volume}{1},
  \bibinfo{number}{1} (\bibinfo{year}{2015}),
  \bibinfo{pages}{2056305115580479}.
\newblock


\bibitem[Gillespie(2017)]%
        {gillespie2017governance}
\bibfield{author}{\bibinfo{person}{Tarleton Gillespie}.}
  \bibinfo{year}{2017}\natexlab{}.
\newblock \showarticletitle{Governance of and by platforms}.
\newblock \bibinfo{journal}{\emph{Sage handbook of social media. London: Sage}}
  (\bibinfo{year}{2017}).
\newblock


\bibitem[Gillespie(2018)]%
        {gillespie2018custodians}
\bibfield{author}{\bibinfo{person}{Tarleton Gillespie}.}
  \bibinfo{year}{2018}\natexlab{}.
\newblock \bibinfo{booktitle}{\emph{Custodians of the Internet: Platforms,
  content moderation, and the hidden decisions that shape social media}}.
\newblock \bibinfo{publisher}{Yale University Press}.
\newblock


\bibitem[Gillespie(2022)]%
        {gillespie2022not}
\bibfield{author}{\bibinfo{person}{Tarleton Gillespie}.}
  \bibinfo{year}{2022}\natexlab{}.
\newblock \showarticletitle{Do Not Recommend? Reduction as a Form of Content
  Moderation}.
\newblock \bibinfo{journal}{\emph{Social Media+ Society}} \bibinfo{volume}{8},
  \bibinfo{number}{3} (\bibinfo{year}{2022}),
  \bibinfo{pages}{20563051221117552}.
\newblock


\bibitem[Godwin(2003)]%
        {godwin2003cyber}
\bibfield{author}{\bibinfo{person}{Mike Godwin}.}
  \bibinfo{year}{2003}\natexlab{}.
\newblock \bibinfo{booktitle}{\emph{Cyber rights: Defending free speech in the
  digital age}}.
\newblock \bibinfo{publisher}{MIT press}.
\newblock


\bibitem[Goldman(2018)]%
        {goldman2018complicated}
\bibfield{author}{\bibinfo{person}{Eric Goldman}.}
  \bibinfo{year}{2018}\natexlab{}.
\newblock \showarticletitle{The complicated story of FOSTA and section 230}.
\newblock \bibinfo{journal}{\emph{First Amend. L. Rev.}}  \bibinfo{volume}{17}
  (\bibinfo{year}{2018}), \bibinfo{pages}{279}.
\newblock


\bibitem[Golhar and Stamm(1991)]%
        {golhar1991just}
\bibfield{author}{\bibinfo{person}{Damodar~Y Golhar} {and}
  \bibinfo{person}{Carol~Lee Stamm}.} \bibinfo{year}{1991}\natexlab{}.
\newblock \showarticletitle{The just-in-time philosophy: a literature review}.
\newblock \bibinfo{journal}{\emph{The International Journal of Production
  Research}} \bibinfo{volume}{29}, \bibinfo{number}{4} (\bibinfo{year}{1991}),
  \bibinfo{pages}{657--676}.
\newblock


\bibitem[Graeber(2023)]%
        {bluesky}
\bibfield{author}{\bibinfo{person}{Jay Graeber}.}
  \bibinfo{year}{2023}\natexlab{}.
\newblock \bibinfo{title}{Composable Moderation}.
\newblock
\newblock
\urldef\tempurl%
\url{https://blueskyweb.xyz/blog/4-13-2023-moderation}
\showURL{%
\tempurl}


\bibitem[Grimmelmann(2015)]%
        {grimmelmann2015virtues}
\bibfield{author}{\bibinfo{person}{James Grimmelmann}.}
  \bibinfo{year}{2015}\natexlab{}.
\newblock \showarticletitle{The virtues of moderation}.
\newblock \bibinfo{journal}{\emph{Yale JL \& Tech.}}  \bibinfo{volume}{17}
  (\bibinfo{year}{2015}), \bibinfo{pages}{42}.
\newblock


\bibitem[Harper et~al\mbox{.}(2015)]%
        {harper2015putting}
\bibfield{author}{\bibinfo{person}{F~Maxwell Harper}, \bibinfo{person}{Funing
  Xu}, \bibinfo{person}{Harmanpreet Kaur}, \bibinfo{person}{Kyle Condiff},
  \bibinfo{person}{Shuo Chang}, {and} \bibinfo{person}{Loren Terveen}.}
  \bibinfo{year}{2015}\natexlab{}.
\newblock \showarticletitle{Putting users in control of their recommendations}.
  In \bibinfo{booktitle}{\emph{Proceedings of the 9th ACM Conference on
  Recommender Systems}}. \bibinfo{pages}{3--10}.
\newblock


\bibitem[Hatmaker(2021)]%
        {hatmaker_2021}
\bibfield{author}{\bibinfo{person}{Taylor Hatmaker}.}
  \bibinfo{year}{2021}\natexlab{}.
\newblock \bibinfo{title}{Discord buys Sentropy, which makes AI software that
  fights online harassment}.
\newblock
\newblock
\urldef\tempurl%
\url{https://techcrunch.com/2021/07/13/discord-buys-sentropy/}
\showURL{%
\tempurl}


\bibitem[He et~al\mbox{.}(2016)]%
        {he2016interactive}
\bibfield{author}{\bibinfo{person}{Chen He}, \bibinfo{person}{Denis Parra},
  {and} \bibinfo{person}{Katrien Verbert}.} \bibinfo{year}{2016}\natexlab{}.
\newblock \showarticletitle{Interactive recommender systems: A survey of the
  state of the art and future research challenges and opportunities}.
\newblock \bibinfo{journal}{\emph{Expert Systems with Applications}}
  \bibinfo{volume}{56} (\bibinfo{year}{2016}), \bibinfo{pages}{9--27}.
\newblock


\bibitem[Herlocker et~al\mbox{.}(2000)]%
        {herlocker2000explaining}
\bibfield{author}{\bibinfo{person}{Jonathan~L Herlocker},
  \bibinfo{person}{Joseph~A Konstan}, {and} \bibinfo{person}{John Riedl}.}
  \bibinfo{year}{2000}\natexlab{}.
\newblock \showarticletitle{Explaining collaborative filtering
  recommendations}. In \bibinfo{booktitle}{\emph{Proceedings of the 2000 ACM
  conference on Computer supported cooperative work}}. ACM,
  \bibinfo{pages}{241--250}.
\newblock


\bibitem[Hsu et~al\mbox{.}(2020)]%
        {hsu2020awareness}
\bibfield{author}{\bibinfo{person}{Silas Hsu}, \bibinfo{person}{Kristen
  Vaccaro}, \bibinfo{person}{Yin Yue}, \bibinfo{person}{Aimee Rickman}, {and}
  \bibinfo{person}{Karrie Karahalios}.} \bibinfo{year}{2020}\natexlab{}.
\newblock \bibinfo{booktitle}{\emph{Awareness, Navigation, and Use of Feed
  Control Settings Online}}.
\newblock \bibinfo{publisher}{Association for Computing Machinery},
  \bibinfo{address}{New York, NY, USA}, \bibinfo{pages}{1–13}.
\newblock
\showISBNx{9781450367080}
\urldef\tempurl%
\url{https://doi.org/10.1145/3313831.3376583}
\showURL{%
\tempurl}


\bibitem[Hutchinson et~al\mbox{.}(2003)]%
        {hutchinson2003technology}
\bibfield{author}{\bibinfo{person}{Hilary Hutchinson}, \bibinfo{person}{Wendy
  Mackay}, \bibinfo{person}{Bo Westerlund}, \bibinfo{person}{Benjamin~B
  Bederson}, \bibinfo{person}{Allison Druin}, \bibinfo{person}{Catherine
  Plaisant}, \bibinfo{person}{Michel Beaudouin-Lafon},
  \bibinfo{person}{St{\'e}phane Conversy}, \bibinfo{person}{Helen Evans},
  \bibinfo{person}{Heiko Hansen}, {et~al\mbox{.}}}
  \bibinfo{year}{2003}\natexlab{}.
\newblock \showarticletitle{Technology probes: inspiring design for and with
  families}. In \bibinfo{booktitle}{\emph{Proceedings of the SIGCHI conference
  on Human factors in computing systems}}. \bibinfo{pages}{17--24}.
\newblock


\bibitem[Instagram(2021)]%
        {instagram}
\bibfield{author}{\bibinfo{person}{Instagram}.}
  \bibinfo{year}{2021}\natexlab{}.
\newblock \bibinfo{title}{Introducing Sensitive Content Control}.
\newblock
\newblock
\urldef\tempurl%
\url{https://about.instagram.com/blog/announcements/introducing-sensitive-content-control}
\showURL{%
\tempurl}


\bibitem[Jannach et~al\mbox{.}(2019)]%
        {jannach2019explanations}
\bibfield{author}{\bibinfo{person}{Dietmar Jannach}, \bibinfo{person}{Michael
  Jugovac}, {and} \bibinfo{person}{Ingrid Nunes}.}
  \bibinfo{year}{2019}\natexlab{}.
\newblock \showarticletitle{Explanations and user control in recommender
  systems}. In \bibinfo{booktitle}{\emph{Proceedings of the 23rd International
  Workshop on Personalization and Recommendation on the Web and Beyond}}.
  \bibinfo{pages}{31--31}.
\newblock


\bibitem[Jannach et~al\mbox{.}(2016)]%
        {jannach2016user}
\bibfield{author}{\bibinfo{person}{Dietmar Jannach}, \bibinfo{person}{Sidra
  Naveed}, {and} \bibinfo{person}{Michael Jugovac}.}
  \bibinfo{year}{2016}\natexlab{}.
\newblock \showarticletitle{User control in recommender systems: Overview and
  interaction challenges}. In \bibinfo{booktitle}{\emph{International
  Conference on Electronic Commerce and Web Technologies}}. Springer,
  \bibinfo{pages}{21--33}.
\newblock


\bibitem[Jhaver et~al\mbox{.}(2019a)]%
        {jhaver2019survey}
\bibfield{author}{\bibinfo{person}{Shagun Jhaver},
  \bibinfo{person}{Darren~Scott Appling}, \bibinfo{person}{Eric Gilbert}, {and}
  \bibinfo{person}{Amy Bruckman}.} \bibinfo{year}{2019}\natexlab{a}.
\newblock \showarticletitle{“Did You Suspect the Post Would Be Removed?”:
  Understanding User Reactions to Content Removals on Reddit}.
\newblock \bibinfo{journal}{\emph{Proc. ACM Hum.-Comput. Interact.}}
  \bibinfo{volume}{3}, \bibinfo{number}{CSCW}, Article \bibinfo{articleno}{192}
  (\bibinfo{date}{Nov.} \bibinfo{year}{2019}), \bibinfo{numpages}{33}~pages.
\newblock
\urldef\tempurl%
\url{https://doi.org/10.1145/3359294}
\showDOI{\tempurl}


\bibitem[Jhaver et~al\mbox{.}(2019b)]%
        {jhaver2019automated}
\bibfield{author}{\bibinfo{person}{Shagun Jhaver}, \bibinfo{person}{Iris
  Birman}, \bibinfo{person}{Eric Gilbert}, {and} \bibinfo{person}{Amy
  Bruckman}.} \bibinfo{year}{2019}\natexlab{b}.
\newblock \showarticletitle{Human-Machine Collaboration for Content Regulation:
  The Case of Reddit Automoderator}.
\newblock \bibinfo{journal}{\emph{ACM Trans. Comput.-Hum. Interact.}}
  \bibinfo{volume}{26}, \bibinfo{number}{5}, Article \bibinfo{articleno}{31}
  (\bibinfo{date}{July} \bibinfo{year}{2019}), \bibinfo{numpages}{35}~pages.
\newblock
\showISSN{1073-0516}
\urldef\tempurl%
\url{https://doi.org/10.1145/3338243}
\showDOI{\tempurl}


\bibitem[Jhaver et~al\mbox{.}(2018a)]%
        {jhaver2018view}
\bibfield{author}{\bibinfo{person}{Shagun Jhaver}, \bibinfo{person}{Larry
  Chan}, {and} \bibinfo{person}{Amy Bruckman}.}
  \bibinfo{year}{2018}\natexlab{a}.
\newblock \showarticletitle{{The View from the Other Side: The Border Between
  Controversial Speech and Harassment on Kotaku in Action}}.
\newblock \bibinfo{journal}{\emph{First Monday}} \bibinfo{volume}{23},
  \bibinfo{number}{2} (\bibinfo{year}{2018}).
\newblock
\urldef\tempurl%
\url{http://firstmonday.org/ojs/index.php/fm/article/view/8232}
\showURL{%
\tempurl}


\bibitem[Jhaver et~al\mbox{.}(2022)]%
        {jhaver2022filterbuddy}
\bibfield{author}{\bibinfo{person}{Shagun Jhaver}, \bibinfo{person}{Quan~Ze
  Chen}, \bibinfo{person}{Detlef Knauss}, {and} \bibinfo{person}{Amy~X.
  Zhang}.} \bibinfo{year}{2022}\natexlab{}.
\newblock \showarticletitle{Designing Word Filter Tools for Creator-Led Comment
  Moderation}. In \bibinfo{booktitle}{\emph{Proceedings of the 2022 CHI
  Conference on Human Factors in Computing Systems}} (New Orleans, LA, USA)
  \emph{(\bibinfo{series}{CHI '22})}. \bibinfo{publisher}{Association for
  Computing Machinery}, \bibinfo{address}{New York, NY, USA}, Article
  \bibinfo{articleno}{205}, \bibinfo{numpages}{21}~pages.
\newblock
\showISBNx{9781450391573}
\urldef\tempurl%
\url{https://doi.org/10.1145/3491102.3517505}
\showDOI{\tempurl}


\bibitem[Jhaver et~al\mbox{.}(2018b)]%
        {jhaver2018blocklists}
\bibfield{author}{\bibinfo{person}{Shagun Jhaver}, \bibinfo{person}{Sucheta
  Ghoshal}, \bibinfo{person}{Amy Bruckman}, {and} \bibinfo{person}{Eric
  Gilbert}.} \bibinfo{year}{2018}\natexlab{b}.
\newblock \showarticletitle{Online Harassment and Content Moderation: The Case
  of Blocklists}.
\newblock \bibinfo{journal}{\emph{ACM Trans. Comput.-Hum. Interact.}}
  \bibinfo{volume}{25}, \bibinfo{number}{2}, Article \bibinfo{articleno}{12}
  (\bibinfo{date}{March} \bibinfo{year}{2018}), \bibinfo{numpages}{33}~pages.
\newblock
\showISSN{1073-0516}
\urldef\tempurl%
\url{https://doi.org/10.1145/3185593}
\showDOI{\tempurl}


\bibitem[Jhaver et~al\mbox{.}(2018c)]%
        {jhaver2018airbnb}
\bibfield{author}{\bibinfo{person}{Shagun Jhaver}, \bibinfo{person}{Yoni
  Karpfen}, {and} \bibinfo{person}{Judd Antin}.}
  \bibinfo{year}{2018}\natexlab{c}.
\newblock \showarticletitle{{Algorithmic Anxiety and Coping Strategies of
  Airbnb Hosts}}.
\newblock \bibinfo{journal}{\emph{Proceedings of the 35th Annual ACM Conference
  on Human Factors in Computing Systems}} (\bibinfo{year}{2018}).
\newblock


\bibitem[Jhaver and Zhang(2023)]%
        {jhaver_zhang_2022}
\bibfield{author}{\bibinfo{person}{Shagun Jhaver} {and} \bibinfo{person}{Amy
  Zhang}.} \bibinfo{year}{2023}\natexlab{}.
\newblock \bibinfo{title}{Do Users Want Platform Moderation or Individual
  Control? Examining the Role of Third-Person Effects and Free Speech Support
  in Shaping Moderation Preferences}.  (\bibinfo{year}{2023}).
\newblock
\newblock
\shownote{In Preparation}.


\bibitem[Jiang et~al\mbox{.}(2022)]%
        {jiang2022trade}
\bibfield{author}{\bibinfo{person}{Jialun~Aaron Jiang}, \bibinfo{person}{Peipei
  Nie}, \bibinfo{person}{Jed~R Brubaker}, {and} \bibinfo{person}{Casey
  Fiesler}.} \bibinfo{year}{2022}\natexlab{}.
\newblock \showarticletitle{A Trade-off-centered Framework of Content
  Moderation}.
\newblock \bibinfo{journal}{\emph{arXiv preprint arXiv:2206.03450}}
  (\bibinfo{year}{2022}).
\newblock


\bibitem[Jiang et~al\mbox{.}(2021)]%
        {jiang2021understanding}
\bibfield{author}{\bibinfo{person}{Jialun~Aaron Jiang},
  \bibinfo{person}{Morgan~Klaus Scheuerman}, \bibinfo{person}{Casey Fiesler},
  {and} \bibinfo{person}{Jed~R Brubaker}.} \bibinfo{year}{2021}\natexlab{}.
\newblock \showarticletitle{Understanding international perceptions of the
  severity of harmful content online}.
\newblock \bibinfo{journal}{\emph{PloS one}} \bibinfo{volume}{16},
  \bibinfo{number}{8} (\bibinfo{year}{2021}), \bibinfo{pages}{e0256762}.
\newblock


\bibitem[Jin et~al\mbox{.}(2017)]%
        {jin2017different}
\bibfield{author}{\bibinfo{person}{Yucheng Jin}, \bibinfo{person}{Bruno
  Cardoso}, {and} \bibinfo{person}{Katrien Verbert}.}
  \bibinfo{year}{2017}\natexlab{}.
\newblock \showarticletitle{How do different levels of user control affect
  cognitive load and acceptance of recommendations?}. In
  \bibinfo{booktitle}{\emph{Jin, Y., Cardoso, B. and Verbert, K., 2017, August.
  How do different levels of user control affect cognitive load and acceptance
  of recommendations?. In Proceedings of the 4th Joint Workshop on Interfaces
  and Human Decision Making for Recommender Systems co-located with ACM
  Conference on Recommender Systems (RecSys 2017)}},
  Vol.~\bibinfo{volume}{1884}. CEUR Workshop Proceedings,
  \bibinfo{pages}{35--42}.
\newblock


\bibitem[Kizilcec(2016)]%
        {kizilcec2016much}
\bibfield{author}{\bibinfo{person}{Ren{\'e}~F Kizilcec}.}
  \bibinfo{year}{2016}\natexlab{}.
\newblock \showarticletitle{How much information?: Effects of transparency on
  trust in an algorithmic interface}. In \bibinfo{booktitle}{\emph{Proceedings
  of the 2016 CHI Conference on Human Factors in Computing Systems}}. ACM,
  \bibinfo{pages}{2390--2395}.
\newblock


\bibitem[Klonick(2017)]%
        {klonick2017}
\bibfield{author}{\bibinfo{person}{Kate Klonick}.}
  \bibinfo{year}{2017}\natexlab{}.
\newblock \showarticletitle{The new governors: The people, rules, and processes
  governing online speech}.
\newblock \bibinfo{journal}{\emph{Harv. L. Rev.}}  \bibinfo{volume}{131}
  (\bibinfo{year}{2017}), \bibinfo{pages}{1598}.
\newblock


\bibitem[Knijnenburg et~al\mbox{.}(2012)]%
        {knijnenburg2012inspectability}
\bibfield{author}{\bibinfo{person}{Bart~P Knijnenburg},
  \bibinfo{person}{Svetlin Bostandjiev}, \bibinfo{person}{John O'Donovan},
  {and} \bibinfo{person}{Alfred Kobsa}.} \bibinfo{year}{2012}\natexlab{}.
\newblock \showarticletitle{Inspectability and control in social recommenders}.
  In \bibinfo{booktitle}{\emph{Proceedings of the sixth ACM conference on
  Recommender systems}}. \bibinfo{pages}{43--50}.
\newblock


\bibitem[Konstan and Riedl(2012)]%
        {konstan2012recommender}
\bibfield{author}{\bibinfo{person}{Joseph~A Konstan} {and}
  \bibinfo{person}{John Riedl}.} \bibinfo{year}{2012}\natexlab{}.
\newblock \showarticletitle{Recommender systems: from algorithms to user
  experience}.
\newblock \bibinfo{journal}{\emph{User modeling and user-adapted interaction}}
  \bibinfo{volume}{22}, \bibinfo{number}{1} (\bibinfo{year}{2012}),
  \bibinfo{pages}{101--123}.
\newblock


\bibitem[Kou and Gui(2021)]%
        {kou2021flag}
\bibfield{author}{\bibinfo{person}{Yubo Kou} {and} \bibinfo{person}{Xinning
  Gui}.} \bibinfo{year}{2021}\natexlab{}.
\newblock \showarticletitle{Flag and Flaggability in Automated Moderation: The
  Case of Reporting Toxic Behavior in an Online Game Community}. In
  \bibinfo{booktitle}{\emph{Proceedings of the 2021 CHI Conference on Human
  Factors in Computing Systems}} (Yokohama, Japan) \emph{(\bibinfo{series}{CHI
  '21})}. \bibinfo{publisher}{Association for Computing Machinery},
  \bibinfo{address}{New York, NY, USA}, Article \bibinfo{articleno}{437},
  \bibinfo{numpages}{12}~pages.
\newblock
\showISBNx{9781450380966}
\urldef\tempurl%
\url{https://doi.org/10.1145/3411764.3445279}
\showDOI{\tempurl}


\bibitem[Kumar et~al\mbox{.}(2021)]%
        {kumar2021designing}
\bibfield{author}{\bibinfo{person}{Deepak Kumar}, \bibinfo{person}{Patrick~Gage
  Kelley}, \bibinfo{person}{Sunny Consolvo}, \bibinfo{person}{Joshua Mason},
  \bibinfo{person}{Elie Bursztein}, \bibinfo{person}{Zakir Durumeric},
  \bibinfo{person}{Kurt Thomas}, {and} \bibinfo{person}{Michael Bailey}.}
  \bibinfo{year}{2021}\natexlab{}.
\newblock \showarticletitle{Designing Toxic Content Classification for a
  Diversity of Perspectives}.
\newblock \bibinfo{journal}{\emph{arXiv preprint arXiv:2106.04511}}
  (\bibinfo{year}{2021}).
\newblock


\bibitem[Lee(2018)]%
        {lee2018understanding}
\bibfield{author}{\bibinfo{person}{Min~Kyung Lee}.}
  \bibinfo{year}{2018}\natexlab{}.
\newblock \showarticletitle{Understanding perception of algorithmic decisions:
  Fairness, trust, and emotion in response to algorithmic management}.
\newblock \bibinfo{journal}{\emph{Big Data \& Society}} \bibinfo{volume}{5},
  \bibinfo{number}{1} (\bibinfo{year}{2018}),
  \bibinfo{pages}{2053951718756684}.
\newblock


\bibitem[Leetaru(2019)]%
        {leetaru_2019}
\bibfield{author}{\bibinfo{person}{Kalev Leetaru}.}
  \bibinfo{year}{2019}\natexlab{}.
\newblock \bibinfo{title}{Could personalized content moderation be the future
  of healthy social media?}
\newblock
\newblock
\urldef\tempurl%
\url{https://www.forbes.com/sites/kalevleetaru/2019/07/28/could-personalized-content-moderation-be-the-future-of-healthy-social-media/}
\showURL{%
\tempurl}


\bibitem[Lessig(2009)]%
        {lessig2009code}
\bibfield{author}{\bibinfo{person}{Lawrence Lessig}.}
  \bibinfo{year}{2009}\natexlab{}.
\newblock \bibinfo{booktitle}{\emph{Code: And other laws of cyberspace}}.
\newblock \bibinfo{publisher}{ReadHowYouWant. com}.
\newblock


\bibitem[Lewis(1982)]%
        {lewis1982using}
\bibfield{author}{\bibinfo{person}{Clayton Lewis}.}
  \bibinfo{year}{1982}\natexlab{}.
\newblock \bibinfo{booktitle}{\emph{Using the" thinking-aloud" method in
  cognitive interface design}}.
\newblock \bibinfo{publisher}{IBM TJ Watson Research Center Yorktown Heights,
  NY}.
\newblock


\bibitem[Litman(1999)]%
        {litman1999electronic}
\bibfield{author}{\bibinfo{person}{Jessica Litman}.}
  \bibinfo{year}{1999}\natexlab{}.
\newblock \showarticletitle{Electronic commerce and free speech}.
\newblock \bibinfo{journal}{\emph{Ethics and Information Technology}}
  \bibinfo{volume}{1}, \bibinfo{number}{3} (\bibinfo{year}{1999}),
  \bibinfo{pages}{213--225}.
\newblock


\bibitem[MacKinnon et~al\mbox{.}(2015)]%
        {mackinnon2015fostering}
\bibfield{author}{\bibinfo{person}{Rebecca MacKinnon}, \bibinfo{person}{Elonnai
  Hickok}, \bibinfo{person}{Allon Bar}, {and} \bibinfo{person}{Hae-in Lim}.}
  \bibinfo{year}{2015}\natexlab{}.
\newblock \bibinfo{booktitle}{\emph{Fostering freedom online: The role of
  internet intermediaries}}.
\newblock \bibinfo{publisher}{UNESCO Publishing}.
\newblock


\bibitem[Mahar et~al\mbox{.}(2018)]%
        {mahar2018squadbox}
\bibfield{author}{\bibinfo{person}{Kaitlin Mahar}, \bibinfo{person}{Amy~X
  Zhang}, {and} \bibinfo{person}{David Karger}.}
  \bibinfo{year}{2018}\natexlab{}.
\newblock \showarticletitle{Squadbox: A tool to combat email harassment using
  friendsourced moderation}. In \bibinfo{booktitle}{\emph{Proceedings of the
  2018 CHI Conference on Human Factors in Computing Systems}}.
  \bibinfo{pages}{1--13}.
\newblock


\bibitem[Mahmood and Ricci(2007)]%
        {mahmood2007learning}
\bibfield{author}{\bibinfo{person}{Tariq Mahmood} {and}
  \bibinfo{person}{Francesco Ricci}.} \bibinfo{year}{2007}\natexlab{}.
\newblock \showarticletitle{Learning and adaptivity in interactive recommender
  systems}. In \bibinfo{booktitle}{\emph{Proceedings of the ninth international
  conference on Electronic commerce}}. \bibinfo{pages}{75--84}.
\newblock


\bibitem[Mallari et~al\mbox{.}(2021)]%
        {mallari2021understanding}
\bibfield{author}{\bibinfo{person}{Keri Mallari}, \bibinfo{person}{Spencer
  Williams}, {and} \bibinfo{person}{Gary Hsieh}.}
  \bibinfo{year}{2021}\natexlab{}.
\newblock \showarticletitle{Understanding Analytics Needs of Video Game
  Streamers}. In \bibinfo{booktitle}{\emph{Proceedings of the 2021 CHI
  Conference on Human Factors in Computing Systems}}. \bibinfo{pages}{1--12}.
\newblock


\bibitem[Masnick(2019)]%
        {masnick_2019}
\bibfield{author}{\bibinfo{person}{Mike Masnick}.}
  \bibinfo{year}{2019}\natexlab{}.
\newblock \bibinfo{title}{Protocols, Not Platforms: A Technological Approach to
  Free Speech}.
\newblock
\newblock
\urldef\tempurl%
\url{https://knightcolumbia.org/content/protocols-not-platforms-a-technological-approach-to-free-speech}
\showURL{%
\tempurl}


\bibitem[Matias(2016)]%
        {matias2016civic}
\bibfield{author}{\bibinfo{person}{Nathan~J. Matias}.}
  \bibinfo{year}{2016}\natexlab{}.
\newblock \showarticletitle{The Civic Labor of Online Moderators}. In
  \bibinfo{booktitle}{\emph{Internet Politics and Policy conference}} (Oxford,
  United Kingdom). \bibinfo{address}{Oxford, United Kingdom}.
\newblock


\bibitem[McGillicuddy et~al\mbox{.}(2016)]%
        {mcgillicuddy2016}
\bibfield{author}{\bibinfo{person}{Aiden McGillicuddy},
  \bibinfo{person}{Jean-Gregoire Bernard}, {and} \bibinfo{person}{Jocelyn
  Cranefield}.} \bibinfo{year}{2016}\natexlab{}.
\newblock \showarticletitle{{Controlling Bad Behavior in Online Communities: An
  Examination of Moderation Work}}.
\newblock \bibinfo{journal}{\emph{ICIS 2016 Proceedings}} (\bibinfo{date}{dec}
  \bibinfo{year}{2016}).
\newblock
\urldef\tempurl%
\url{http://aisel.aisnet.org/icis2016/SocialMedia/Presentations/23}
\showURL{%
\tempurl}


\bibitem[Merriam(2002)]%
        {merriam2002}
\bibfield{author}{\bibinfo{person}{Sharan~B Merriam}.}
  \bibinfo{year}{2002}\natexlab{}.
\newblock \showarticletitle{{Introduction to Qualitative Research}}.
\newblock \bibinfo{journal}{\emph{Qualitative research in practice: Examples
  for discussion and analysis}}  \bibinfo{volume}{1} (\bibinfo{year}{2002}).
\newblock


\bibitem[Morrow et~al\mbox{.}(2021)]%
        {morrow2021emerging}
\bibfield{author}{\bibinfo{person}{Garrett Morrow}, \bibinfo{person}{Briony
  Swire-Thompson}, \bibinfo{person}{Jessica~Montgomery Polny},
  \bibinfo{person}{Matthew Kopec}, {and} \bibinfo{person}{John~P Wihbey}.}
  \bibinfo{year}{2021}\natexlab{}.
\newblock \showarticletitle{The emerging science of content labeling:
  Contextualizing social media content moderation}.
\newblock \bibinfo{journal}{\emph{Journal of the Association for Information
  Science and Technology}} (\bibinfo{year}{2021}).
\newblock


\bibitem[Narayanan(2022)]%
        {narayanan2022Tiktok}
\bibfield{author}{\bibinfo{person}{Arvind Narayanan}.}
  \bibinfo{year}{2022}\natexlab{}.
\newblock \bibinfo{title}{How to train your TikTok}.
\newblock
\newblock
\urldef\tempurl%
\url{https://knightcolumbia.org/blog/how-to-train-your-tiktok}
\showURL{%
\tempurl}


\bibitem[Obar and Wildman(2015)]%
        {obar2015social}
\bibfield{author}{\bibinfo{person}{Jonathan~A Obar} {and}
  \bibinfo{person}{Steven~S Wildman}.} \bibinfo{year}{2015}\natexlab{}.
\newblock \showarticletitle{Social media definition and the governance
  challenge-an introduction to the special issue}.
\newblock \bibinfo{journal}{\emph{Obar, JA and Wildman, S.(2015). Social media
  definition and the governance challenge: An introduction to the special
  issue. Telecommunications policy}} \bibinfo{volume}{39}, \bibinfo{number}{9}
  (\bibinfo{year}{2015}), \bibinfo{pages}{745--750}.
\newblock


\bibitem[Parra and Brusilovsky(2015)]%
        {parra2015user}
\bibfield{author}{\bibinfo{person}{Denis Parra} {and} \bibinfo{person}{Peter
  Brusilovsky}.} \bibinfo{year}{2015}\natexlab{}.
\newblock \showarticletitle{User-controllable personalization: A case study
  with SetFusion}.
\newblock \bibinfo{journal}{\emph{International Journal of Human-Computer
  Studies}}  \bibinfo{volume}{78} (\bibinfo{year}{2015}),
  \bibinfo{pages}{43--67}.
\newblock


\bibitem[Party(2022)]%
        {blockparty}
\bibfield{author}{\bibinfo{person}{Block Party}.}
  \bibinfo{year}{2022}\natexlab{}.
\newblock \bibinfo{title}{Block Party}.
\newblock
\newblock
\urldef\tempurl%
\url{https://www.blockpartyapp.com}
\showURL{%
\tempurl}


\bibitem[Pasquale(2015)]%
        {pasquale2015black}
\bibfield{author}{\bibinfo{person}{Frank Pasquale}.}
  \bibinfo{year}{2015}\natexlab{}.
\newblock \bibinfo{booktitle}{\emph{The black box society: The secret
  algorithms that control money and information}}.
\newblock \bibinfo{publisher}{Harvard University Press}.
\newblock


\bibitem[Pater et~al\mbox{.}(2016)]%
        {pater2016}
\bibfield{author}{\bibinfo{person}{Jessica~A. Pater}, \bibinfo{person}{Moon~K.
  Kim}, \bibinfo{person}{Elizabeth~D. Mynatt}, {and} \bibinfo{person}{Casey
  Fiesler}.} \bibinfo{year}{2016}\natexlab{}.
\newblock \showarticletitle{Characterizations of Online Harassment: Comparing
  Policies Across Social Media Platforms}. In
  \bibinfo{booktitle}{\emph{Proceedings of the 19th International Conference on
  Supporting Group Work}} (Sanibel Island, Florida, USA)
  \emph{(\bibinfo{series}{GROUP '16})}. \bibinfo{publisher}{ACM},
  \bibinfo{address}{New York, NY, USA}, \bibinfo{pages}{369--374}.
\newblock
\showISBNx{978-1-4503-4276-6}
\urldef\tempurl%
\url{https://doi.org/10.1145/2957276.2957297}
\showDOI{\tempurl}


\bibitem[Porter(2021)]%
        {porter_2021}
\bibfield{author}{\bibinfo{person}{Jon Porter}.}
  \bibinfo{year}{2021}\natexlab{}.
\newblock \bibinfo{title}{Today I learned about Intel's AI sliders that filter
  online gaming abuse}.
\newblock
\newblock
\urldef\tempurl%
\url{https://www.theverge.com/2021/4/8/22373290/intel-bleep-ai-powered-abuse-toxicity-gaming-filters}
\showURL{%
\tempurl}


\bibitem[Pu et~al\mbox{.}(2012)]%
        {pu2012evaluating}
\bibfield{author}{\bibinfo{person}{Pearl Pu}, \bibinfo{person}{Li Chen}, {and}
  \bibinfo{person}{Rong Hu}.} \bibinfo{year}{2012}\natexlab{}.
\newblock \showarticletitle{Evaluating recommender systems from the user's
  perspective: survey of the state of the art}.
\newblock \bibinfo{journal}{\emph{User Modeling and User-Adapted Interaction}}
  \bibinfo{volume}{22}, \bibinfo{number}{4} (\bibinfo{year}{2012}),
  \bibinfo{pages}{317--355}.
\newblock


\bibitem[Reid~Chassiakos et~al\mbox{.}(2016)]%
        {reid2016children}
\bibfield{author}{\bibinfo{person}{Yolanda~Linda Reid~Chassiakos},
  \bibinfo{person}{Jenny Radesky}, \bibinfo{person}{Dimitri Christakis},
  \bibinfo{person}{Megan~A Moreno}, \bibinfo{person}{Corinn Cross},
  \bibinfo{person}{David Hill}, \bibinfo{person}{Nusheen Ameenuddin},
  \bibinfo{person}{Jeffrey Hutchinson}, \bibinfo{person}{Alanna Levine},
  \bibinfo{person}{Rhea Boyd}, {et~al\mbox{.}}}
  \bibinfo{year}{2016}\natexlab{}.
\newblock \showarticletitle{Children and adolescents and digital media}.
\newblock \bibinfo{journal}{\emph{Pediatrics}} \bibinfo{volume}{138},
  \bibinfo{number}{5} (\bibinfo{year}{2016}).
\newblock


\bibitem[Roberts(2019)]%
        {roberts2019behind}
\bibfield{author}{\bibinfo{person}{Sarah~T Roberts}.}
  \bibinfo{year}{2019}\natexlab{}.
\newblock \bibinfo{booktitle}{\emph{Behind the Screen: Content Moderation in
  the Shadows of Social Media}}.
\newblock \bibinfo{publisher}{Yale University Press}.
\newblock


\bibitem[Roy et~al\mbox{.}(2019)]%
        {Roy2019AutomationAI}
\bibfield{author}{\bibinfo{person}{Quentin Roy}, \bibinfo{person}{Futian
  Zhang}, {and} \bibinfo{person}{Daniel Vogel}.}
  \bibinfo{year}{2019}\natexlab{}.
\newblock \showarticletitle{Automation Accuracy Is Good, but High
  Controllability May Be Better}.
\newblock \bibinfo{journal}{\emph{Proceedings of the 2019 CHI Conference on
  Human Factors in Computing Systems}} (\bibinfo{year}{2019}).
\newblock


\bibitem[Scheuerman et~al\mbox{.}(2018)]%
        {scheuerman2018safe}
\bibfield{author}{\bibinfo{person}{Morgan~Klaus Scheuerman},
  \bibinfo{person}{Stacy~M. Branham}, {and} \bibinfo{person}{Foad Hamidi}.}
  \bibinfo{year}{2018}\natexlab{}.
\newblock \showarticletitle{Safe Spaces and Safe Places: Unpacking
  Technology-Mediated Experiences of Safety and Harm with Transgender People}.
\newblock \bibinfo{journal}{\emph{Proc. ACM Hum.-Comput. Interact.}}
  \bibinfo{volume}{2}, \bibinfo{number}{CSCW}, Article \bibinfo{articleno}{155}
  (\bibinfo{date}{nov} \bibinfo{year}{2018}), \bibinfo{numpages}{27}~pages.
\newblock
\urldef\tempurl%
\url{https://doi.org/10.1145/3274424}
\showDOI{\tempurl}


\bibitem[Schoenebeck et~al\mbox{.}(2020)]%
        {schoenebeck2020drawing}
\bibfield{author}{\bibinfo{person}{Sarita Schoenebeck},
  \bibinfo{person}{Oliver~L Haimson}, {and} \bibinfo{person}{Lisa Nakamura}.}
  \bibinfo{year}{2020}\natexlab{}.
\newblock \showarticletitle{Drawing from justice theories to support targets of
  online harassment}.
\newblock \bibinfo{journal}{\emph{New Media \& Society}}
  (\bibinfo{year}{2020}).
\newblock
\urldef\tempurl%
\url{https://doi.org/10.1177/1461444820913122}
\showURL{%
\tempurl}


\bibitem[Seering et~al\mbox{.}(2019)]%
        {seering2019moderator}
\bibfield{author}{\bibinfo{person}{Joseph Seering}, \bibinfo{person}{Tony
  Wang}, \bibinfo{person}{Jina Yoon}, {and} \bibinfo{person}{Geoff Kaufman}.}
  \bibinfo{year}{2019}\natexlab{}.
\newblock \showarticletitle{Moderator engagement and community development in
  the age of algorithms}.
\newblock \bibinfo{journal}{\emph{New Media \& Society}}
  (\bibinfo{year}{2019}), \bibinfo{pages}{1461444818821316}.
\newblock


\bibitem[Shah and Sandvig(2008)]%
        {shah2008software}
\bibfield{author}{\bibinfo{person}{Rajiv~C Shah} {and}
  \bibinfo{person}{Christian Sandvig}.} \bibinfo{year}{2008}\natexlab{}.
\newblock \showarticletitle{Software defaults as de facto regulation the case
  of the wireless Internet}.
\newblock \bibinfo{journal}{\emph{Information, Community \& Society}}
  \bibinfo{volume}{11}, \bibinfo{number}{1} (\bibinfo{year}{2008}),
  \bibinfo{pages}{25--46}.
\newblock


\bibitem[Sinha and Swearingen(2002)]%
        {sinha2002role}
\bibfield{author}{\bibinfo{person}{Rashmi Sinha} {and} \bibinfo{person}{Kirsten
  Swearingen}.} \bibinfo{year}{2002}\natexlab{}.
\newblock \showarticletitle{The role of transparency in recommender systems}.
  In \bibinfo{booktitle}{\emph{CHI'02 extended abstracts on Human factors in
  computing systems}}. \bibinfo{pages}{830--831}.
\newblock


\bibitem[Steiger et~al\mbox{.}(2021)]%
        {steiger2021psy}
\bibfield{author}{\bibinfo{person}{Miriah Steiger}, \bibinfo{person}{Timir~J
  Bharucha}, \bibinfo{person}{Sukrit Venkatagiri}, \bibinfo{person}{Martin~J.
  Riedl}, {and} \bibinfo{person}{Matthew Lease}.}
  \bibinfo{year}{2021}\natexlab{}.
\newblock \showarticletitle{The Psychological Well-Being of Content Moderators:
  The Emotional Labor of Commercial Moderation and Avenues for Improving
  Support}. In \bibinfo{booktitle}{\emph{Proceedings of the 2021 CHI Conference
  on Human Factors in Computing Systems}} (Yokohama, Japan)
  \emph{(\bibinfo{series}{CHI '21})}. \bibinfo{publisher}{Association for
  Computing Machinery}, \bibinfo{address}{New York, NY, USA}, Article
  \bibinfo{articleno}{341}, \bibinfo{numpages}{14}~pages.
\newblock
\showISBNx{9781450380966}
\urldef\tempurl%
\url{https://doi.org/10.1145/3411764.3445092}
\showDOI{\tempurl}


\bibitem[Stray et~al\mbox{.}(2022)]%
        {stray2022building}
\bibfield{author}{\bibinfo{person}{Jonathan Stray}, \bibinfo{person}{Alon
  Halevy}, \bibinfo{person}{Parisa Assar}, \bibinfo{person}{Dylan
  Hadfield-Menell}, \bibinfo{person}{Craig Boutilier}, \bibinfo{person}{Amar
  Ashar}, \bibinfo{person}{Lex Beattie}, \bibinfo{person}{Michael Ekstrand},
  \bibinfo{person}{Claire Leibowicz}, \bibinfo{person}{Connie~Moon Sehat},
  {et~al\mbox{.}}} \bibinfo{year}{2022}\natexlab{}.
\newblock \showarticletitle{Building Human Values into Recommender Systems: An
  Interdisciplinary Synthesis}.
\newblock \bibinfo{journal}{\emph{arXiv preprint arXiv:2207.10192}}
  (\bibinfo{year}{2022}).
\newblock


\bibitem[Stubbs et~al\mbox{.}(2022)]%
        {stubbs2022investigating}
\bibfield{author}{\bibinfo{person}{Joshua Stubbs}, \bibinfo{person}{Laura
  Nicklin}, \bibinfo{person}{Luke Wilsdon}, {and} \bibinfo{person}{Joanne
  Lloyd}.} \bibinfo{year}{2022}\natexlab{}.
\newblock \showarticletitle{Investigating the experience of viewing extreme
  real-world violence online: naturalistic evidence from an online discussion
  forum}.
\newblock  (\bibinfo{year}{2022}).
\newblock


\bibitem[Suzor et~al\mbox{.}(2019)]%
        {suzor2019we}
\bibfield{author}{\bibinfo{person}{Nicolas~P Suzor},
  \bibinfo{person}{Sarah~Myers West}, \bibinfo{person}{Andrew Quodling}, {and}
  \bibinfo{person}{Jillian York}.} \bibinfo{year}{2019}\natexlab{}.
\newblock \showarticletitle{What Do We Mean When We Talk About Transparency?
  Toward Meaningful Transparency in Commercial Content Moderation}.
\newblock \bibinfo{journal}{\emph{International Journal of Communication}}
  \bibinfo{volume}{13} (\bibinfo{year}{2019}), \bibinfo{pages}{18}.
\newblock


\bibitem[Tait(2008)]%
        {tait2008pornographies}
\bibfield{author}{\bibinfo{person}{Sue Tait}.} \bibinfo{year}{2008}\natexlab{}.
\newblock \showarticletitle{Pornographies of violence? Internet spectatorship
  on body horror}.
\newblock \bibinfo{journal}{\emph{Critical Studies in Media Communication}}
  \bibinfo{volume}{25}, \bibinfo{number}{1} (\bibinfo{year}{2008}),
  \bibinfo{pages}{91--111}.
\newblock


\bibitem[Taylor(2018)]%
        {taylor2018watch}
\bibfield{author}{\bibinfo{person}{TL Taylor}.}
  \bibinfo{year}{2018}\natexlab{}.
\newblock \showarticletitle{Regulating the networked broadcasting frontier}.
\newblock In \bibinfo{booktitle}{\emph{Watch me play: Twitch and the rise of
  game live streaming}}. \bibinfo{publisher}{Princeton University Press},
  Chapter~5.
\newblock


\bibitem[Telecommunications~Union(2021)]%
        {union_2021}
\bibfield{author}{\bibinfo{person}{International Telecommunications~Union}.}
  \bibinfo{year}{2021}\natexlab{}.
\newblock
\newblock
\urldef\tempurl%
\url{https://www.itu.int/itu-d/reports/statistics/facts-figures-2021/}
\showURL{%
\tempurl}


\bibitem[Tintarev and Masthoff(2011)]%
        {tintarev2011designing}
\bibfield{author}{\bibinfo{person}{Nava Tintarev} {and} \bibinfo{person}{Judith
  Masthoff}.} \bibinfo{year}{2011}\natexlab{}.
\newblock \showarticletitle{Designing and evaluating explanations for
  recommender systems}.
\newblock In \bibinfo{booktitle}{\emph{Recommender systems handbook}}.
  \bibinfo{publisher}{Springer}, \bibinfo{pages}{479--510}.
\newblock


\bibitem[Tushnet(2019)]%
        {tushnet2019content}
\bibfield{author}{\bibinfo{person}{Rebecca Tushnet}.}
  \bibinfo{year}{2019}\natexlab{}.
\newblock \showarticletitle{Content moderation in an age of extremes}.
\newblock \bibinfo{journal}{\emph{Case W. Res. JL Tech. \& Internet}}
  \bibinfo{volume}{10} (\bibinfo{year}{2019}), \bibinfo{pages}{1}.
\newblock


\bibitem[Twenge et~al\mbox{.}(2019)]%
        {twenge2019trends}
\bibfield{author}{\bibinfo{person}{Jean~M Twenge}, \bibinfo{person}{Gabrielle~N
  Martin}, {and} \bibinfo{person}{Brian~H Spitzberg}.}
  \bibinfo{year}{2019}\natexlab{}.
\newblock \showarticletitle{Trends in US Adolescents' media use, 1976--2016:
  The rise of digital media, the decline of TV, and the (near) demise of
  print.}
\newblock \bibinfo{journal}{\emph{Psychology of Popular Media Culture}}
  \bibinfo{volume}{8}, \bibinfo{number}{4} (\bibinfo{year}{2019}),
  \bibinfo{pages}{329}.
\newblock


\bibitem[Vaccaro et~al\mbox{.}(2018)]%
        {vaccaro2018illusion}
\bibfield{author}{\bibinfo{person}{Kristen Vaccaro}, \bibinfo{person}{Dylan
  Huang}, \bibinfo{person}{Motahhare Eslami}, \bibinfo{person}{Christian
  Sandvig}, \bibinfo{person}{Kevin Hamilton}, {and} \bibinfo{person}{Karrie
  Karahalios}.} \bibinfo{year}{2018}\natexlab{}.
\newblock \showarticletitle{The illusion of control: Placebo effects of control
  settings}. In \bibinfo{booktitle}{\emph{Proceedings of the 2018 CHI
  Conference on Human Factors in Computing Systems}}. \bibinfo{pages}{1--13}.
\newblock


\bibitem[Vaccaro et~al\mbox{.}(2020)]%
        {vaccaro2020at}
\bibfield{author}{\bibinfo{person}{Kristen Vaccaro}, \bibinfo{person}{Christian
  Sandvig}, {and} \bibinfo{person}{Karrie Karahalios}.}
  \bibinfo{year}{2020}\natexlab{}.
\newblock \showarticletitle{"At the End of the Day Facebook Does What It
  Wants": How Users Experience Contesting Algorithmic Content Moderation}.
\newblock \bibinfo{journal}{\emph{Proc. ACM Hum.-Comput. Interact.}}
  \bibinfo{volume}{4}, \bibinfo{number}{CSCW2}, Article
  \bibinfo{articleno}{167} (\bibinfo{date}{oct} \bibinfo{year}{2020}),
  \bibinfo{numpages}{22}~pages.
\newblock
\urldef\tempurl%
\url{https://doi.org/10.1145/3415238}
\showDOI{\tempurl}


\bibitem[Vaccaro et~al\mbox{.}(2021)]%
        {vaccaro2021cont}
\bibfield{author}{\bibinfo{person}{Kristen Vaccaro}, \bibinfo{person}{Ziang
  Xiao}, \bibinfo{person}{Kevin Hamilton}, {and} \bibinfo{person}{Karrie
  Karahalios}.} \bibinfo{year}{2021}\natexlab{}.
\newblock \showarticletitle{Contestability For Content Moderation}.
\newblock  \bibinfo{volume}{5}, \bibinfo{number}{CSCW2}, Article
  \bibinfo{articleno}{318} (\bibinfo{date}{oct} \bibinfo{year}{2021}),
  \bibinfo{numpages}{28}~pages.
\newblock
\urldef\tempurl%
\url{https://doi.org/10.1145/3476059}
\showDOI{\tempurl}


\bibitem[Van~Dijck(2013)]%
        {van2013culture}
\bibfield{author}{\bibinfo{person}{Jos{\'e} Van~Dijck}.}
  \bibinfo{year}{2013}\natexlab{}.
\newblock \bibinfo{booktitle}{\emph{The culture of connectivity: A critical
  history of social media}}.
\newblock \bibinfo{publisher}{Oxford University Press}.
\newblock


\bibitem[Van~Someren et~al\mbox{.}(1994)]%
        {van1994think}
\bibfield{author}{\bibinfo{person}{Maarten~W Van~Someren},
  \bibinfo{person}{Yvonne~F Barnard}, {and} \bibinfo{person}{Jacobijn~AC
  Sandberg}.} \bibinfo{year}{1994}\natexlab{}.
\newblock \showarticletitle{The think aloud method: a practical approach to
  modelling cognitive}.
\newblock \bibinfo{journal}{\emph{London: AcademicPress}}  \bibinfo{volume}{11}
  (\bibinfo{year}{1994}).
\newblock


\bibitem[Vig et~al\mbox{.}(2009)]%
        {vig2009tagsplanations}
\bibfield{author}{\bibinfo{person}{Jesse Vig}, \bibinfo{person}{Shilad Sen},
  {and} \bibinfo{person}{John Riedl}.} \bibinfo{year}{2009}\natexlab{}.
\newblock \showarticletitle{Tagsplanations: explaining recommendations using
  tags}. In \bibinfo{booktitle}{\emph{Proceedings of the 14th international
  conference on Intelligent user interfaces}}. \bibinfo{pages}{47--56}.
\newblock


\bibitem[Wagner(2013)]%
        {wagner2013governing}
\bibfield{author}{\bibinfo{person}{Ben Wagner}.}
  \bibinfo{year}{2013}\natexlab{}.
\newblock \showarticletitle{Governing internet expression: How public and
  private regulation shape expression governance}.
\newblock \bibinfo{journal}{\emph{Journal of Information Technology \&
  Politics}} \bibinfo{volume}{10}, \bibinfo{number}{4} (\bibinfo{year}{2013}),
  \bibinfo{pages}{389--403}.
\newblock


\bibitem[Weld et~al\mbox{.}(2022)]%
        {weld2022makes}
\bibfield{author}{\bibinfo{person}{Galen Weld}, \bibinfo{person}{Amy~X Zhang},
  {and} \bibinfo{person}{Tim Althoff}.} \bibinfo{year}{2022}\natexlab{}.
\newblock \showarticletitle{What Makes Online Communities `Better'? Measuring
  Values, Consensus, and Conflict across Thousands of Subreddits}. In
  \bibinfo{booktitle}{\emph{Proceedings of the International AAAI Conference on
  Web and Social Media}}, Vol.~\bibinfo{volume}{16}.
  \bibinfo{pages}{1121--1132}.
\newblock


\bibitem[Wellman et~al\mbox{.}(2003)]%
        {wellman2003social}
\bibfield{author}{\bibinfo{person}{Barry Wellman}, \bibinfo{person}{Anabel
  Quan-Haase}, \bibinfo{person}{Jeffrey Boase}, \bibinfo{person}{Wenhong Chen},
  \bibinfo{person}{Keith Hampton}, \bibinfo{person}{Isabel D{\'\i}az}, {and}
  \bibinfo{person}{Kakuko Miyata}.} \bibinfo{year}{2003}\natexlab{}.
\newblock \showarticletitle{The social affordances of the Internet for
  networked individualism}.
\newblock \bibinfo{journal}{\emph{Journal of computer-mediated communication}}
  \bibinfo{volume}{8}, \bibinfo{number}{3} (\bibinfo{year}{2003}),
  \bibinfo{pages}{JCMC834}.
\newblock


\bibitem[West(2018)]%
        {myers2018censored}
\bibfield{author}{\bibinfo{person}{Sarah~Myers West}.}
  \bibinfo{year}{2018}\natexlab{}.
\newblock \showarticletitle{Censored, suspended, shadowbanned: User
  interpretations of content moderation on social media platforms}.
\newblock \bibinfo{journal}{\emph{New Media \& Society}}
  (\bibinfo{year}{2018}).
\newblock


\bibitem[Wilson and Land(2020)]%
        {wilson2020hate}
\bibfield{author}{\bibinfo{person}{Richard~Ashby Wilson} {and}
  \bibinfo{person}{Molly~K Land}.} \bibinfo{year}{2020}\natexlab{}.
\newblock \showarticletitle{Hate speech on social media: Content moderation in
  context}.
\newblock \bibinfo{journal}{\emph{Conn. L. Rev.}}  \bibinfo{volume}{52}
  (\bibinfo{year}{2020}), \bibinfo{pages}{1029}.
\newblock


\bibitem[Wohn(2019)]%
        {wohn2019volunteer}
\bibfield{author}{\bibinfo{person}{Donghee~Yvette Wohn}.}
  \bibinfo{year}{2019}\natexlab{}.
\newblock \showarticletitle{Volunteer Moderators in Twitch Micro Communities:
  How They Get Involved, the Roles They Play, and the Emotional Labor They
  Experience}. In \bibinfo{booktitle}{\emph{Proceedings of the 2019 CHI
  Conference on Human Factors in Computing Systems}}. ACM,
  \bibinfo{pages}{160}.
\newblock


\bibitem[York(2022)]%
        {york2022silicon}
\bibfield{author}{\bibinfo{person}{Jillian~C York}.}
  \bibinfo{year}{2022}\natexlab{}.
\newblock \bibinfo{booktitle}{\emph{Silicon values: The future of free speech
  under surveillance capitalism}}.
\newblock \bibinfo{publisher}{Verso Books}.
\newblock


\bibitem[Zannettou(2021)]%
        {zannettou2021won}
\bibfield{author}{\bibinfo{person}{Savvas Zannettou}.}
  \bibinfo{year}{2021}\natexlab{}.
\newblock \showarticletitle{" I Won the Election!": An Empirical Analysis of
  Soft Moderation Interventions on Twitter.}. In
  \bibinfo{booktitle}{\emph{ICWSM}}. \bibinfo{pages}{865--876}.
\newblock


\bibitem[Zuckerberg(2021)]%
        {zuckerberg2021Building}
\bibfield{author}{\bibinfo{person}{Mark Zuckerberg}.}
  \bibinfo{year}{2021}\natexlab{}.
\newblock \bibinfo{title}{Building Global Community}.
\newblock
\newblock
\urldef\tempurl%
\url{https://www.facebook.com/notes/3707971095882612/}
\showURL{%
\tempurl}


\end{thebibliography}


\end{document}